\documentclass [a4paper,12pt]{article}
\usepackage{amsmath,comment}
\usepackage{amsthm}
\usepackage{amsfonts}
\usepackage{amssymb,float,bm}
\usepackage[pdftex]{graphicx,color} 
\usepackage{latexsym,xcolor,tikz}
\usepackage{latexsym}
\usepackage{graphicx}
\usepackage[font=small,labelfont=]{subfig}
\usepackage{pdfpages}
\usepackage{authblk,svg}
\usepackage{multirow,mathtools}
\usepackage{enumitem}
\graphicspath{{Figures/}}
\usepackage[overload]{empheq}
\DeclareMathOperator{\arcosh}{arcosh}

\DeclareMathOperator{\arsinh}{arsinh}

\title{Rotational flow underlying coupled surface and internal waves. I: Eulerian perspective}
\author[1]{David Henry \thanks{d.henry@ucc.ie}}
\author[2]{Rossen I. Ivanov \thanks{rossen.ivanov@tudublin.ie}}
\author[1]{Zisis N. Sakellaris \thanks{znsakell@hotmail.com}}

\affil[1]{\small{School of Mathematical Sciences, University College Cork, Cork, Ireland}}
\affil[2]{\small{School of Mathematics and Statistics, Technological University Dublin, Grangegorman Lower, Dublin, Ireland}}

\date{}
\begin{document}
\newtheorem{theorem}{Theorem}[section]
\newtheorem{lemma}[theorem]{Lemma}
\newtheorem{proposition}[theorem]{Proposition}
\newtheorem{corollary}[theorem]{Corollary}
\newtheorem{remark}{Remark}[section]
\newenvironment{definition}[1][Definition]{\begin{trivlist}
\item[\hskip \labelsep {\bfseries #1}]}{\end{trivlist}}
\newenvironment{example}[1][Example]{\begin{trivlist}
\item[\hskip \labelsep {\bfseries #1}]}{\end{trivlist}}
\newcommand{\mtp}
{\mathfrak p}
\newcommand{\mtps}
{\mathfrak p^*}
\newcommand{\mtpss}
{\mathfrak p^{**}}
\newcommand{\yu}
{\mathsf{\bar Y_1^*}}
\newcommand{\yuu}
{\mathsf{\tilde Y_1}}
\newcommand{\ym}
{\mathsf{\bar Y_1}}
\newcommand{\yl}
{\mathsf{Y_1^*}}
\newcommand{\ycu}
{\mathcal{\bar Y}_1^*}
\newcommand{\ycm}
{\mathcal{\bar Y}_1}
\newcommand{\ycl}
{\mathcal{Y}_1^*}
\newcommand{\R}
{\mathbb{R}}
\newcommand{\e}
{\mathfrak{e}}
\newcommand{\eps}
{\epsilon}
\newcommand{\Adefa}{\mathcal A}
\newcommand{\lAA}
{\ln\Adefa}
\newcommand{\Adefb}{\tilde{\mathcal A}}
\newcommand{\lA}
{\frac 1 2 \ln\Adefb}
\newcommand{\Fdefa}{\mathcal F}
\newcommand{\Fdefb}{\tilde{\mathcal F}}
\newcommand{\mrl}{\mathfrak{R}}
\newcommand{\mrg}{\tilde{\mathfrak{R}}}
\newcommand{\tr}{\textcolor{red}}
\newcommand{\tb}{\textcolor{blue}}
\newcommand{\tg}{\textcolor{gray}}
\newcommand{\Beq}{\begin{equation}}
\newcommand{\Eeq}{\end{equation}}
\newcommand{\BS}{\begin{subequations}}
\newcommand{\ES}{\end{subequations}}
\newcommand{\Beqn}{\begin{equation*}}
\newcommand{\Eeqn}{\end{equation*}}
\newcommand{\Beqa}{\begin{eqnarray}}
\newcommand{\Eeqa}{\end{eqnarray}}
\newcommand{\Beqan}{\begin{eqnarray*}}
\newcommand{\Eeqan}{\end{eqnarray*}}
\newcommand{\D}{\mathcal D_{\eta}}
\newcommand{\OD}{\overline{\mathcal D_{\eta}}}
\newcommand{\cdef}{\mathfrak{c}}

\definecolor{azure}{rgb}{0.0, 0.5, 1.0} 
\definecolor{bleudefrance}{rgb}{0.16, 0.32, 0.75} 
\definecolor{ballblue}{rgb}{0.13, 0.67, 0.8} 
\definecolor{bblue}{rgb}{0.16, 0.32, 0.74}

\maketitle
\vspace{-1em}

\begin{abstract}
\noindent In this paper we examine the flow generated by coupled surface and internal small-amplitude water waves in a two-fluid layer model, where we take the upper layer to be rotational (constant vorticity) and the lower layer to be irrotational. The presence of vorticity greatly complicates the underlying analysis, yet it generates a rich array of otherwise unobservable phenomena such as the presence of critical layers, and stagnation points, in the fluid interior. We employ a phase-plane analysis to elucidate the qualitative behaviour of streamlines for a range of different coupled-wave, and vorticity, regimes. Although the water waves considered are linear in the fluid dynamics sense, the dynamical systems which govern their motion are nonlinear.
\end{abstract}
\\[.25em]


\mathtoolsset{showonlyrefs=true}

\section{Introduction}
In this paper we examine the flow generated by coupled surface and internal small-amplitude water waves in a two-fluid layer model, where the upper layer is taken to be rotational (constant vorticity), while the lower layer is irrotational. 
This work  significantly advances the recent investigations of fluid kinematics induced by internal and surface wave motion featured in \cite{HV-JDE,HV-AN}, where both fluid layers were assumed to be irrotational.  
Internal water waves  arise where there is a jump in density between fluid layers, which may occur when there are variations in temperature, salinity, or other fluctuations
in the equations of state \cite{Suth}. The two fluid-layer model is applicable when there is an internal interface which demarcates regions with a relatively significant vertical density difference: such an interface is called a pycnocline (or thermocline if the difference in densities  is due to temperature variations). 

The novel feature of this paper is the inclusion of  vorticity in our model. Vorticity greatly complicates the mathematical analysis for the water-wave problem in general \cite{CO}, primarily since the velocity field  can no longer be derived from a potential function. This is particularly germane to the intricacies involved in elucidating the qualitative features of  fluid motion: vorticity has a direct influence on all kinematic considerations.
Incorporating (non-zero) vorticity is necessary for modelling important, and ubiquitous, physical phenomena such as wave-current interactions  \cite{CO,ThomKlop}.  The presence of vorticity can generate a rich array of otherwise unobservable phenomena, such as the presence of critical layers, and stagnation points, in the fluid interior, as we establish in Section \ref{sec-PP}. Note that the fluid model being considered, which consists of  a rotational upper layer overlying an irrotational lower layer, also has direct physical relevance. For instance, wind-generated water waves exhibit a localised layer of vorticity near the surface. In the initial stages of wind generated-waves on the ocean, the wind ``grips'' the surface, producing small-amplitude water waves which subsequently develop in size \cite{Kins}.

Determining the underlying fluid motion generated by a wave propagating on an interface is an intriguing, and broad, area of theoretical research which has witnessed many mathematical advances over the past couple of decades. For periodic waves, a series of work initiated by the paper \cite{CV} used a phase-plane approach to investigate the underlying flow for linear irrotational surface waves on a single fluid layer: see \cite{CO,Con-15,HV-JDE} for additional references. This approach was extended to linear waves with constant vorticity in \cite{EV-JDE}, with further subsequent investigations of the wave kinematics for single layer rotational flow such as \cite{EEV}, among others.  Quite surprisingly, using altogether different methods, it was possible to establish a detailed picture of the underlying flow for fully nonlinear (that is, large amplitude) waves up to, and including, Stokes' wave of extreme height \cite{Con-06,Con-12,Con-15,Hen-06,Hen-08,Lyons-14}. In short, elucidating the fine properties of  individual fluid particle motion has been largely achieved for periodic surface waves.

Recently, in \cite{HV-JDE,HV-AN} a phase-plane approach was employed to obtain a first detailed insight into flow generated by coupled surface and internal waves in the linear regime.
The aim of this article is to significantly extend these results  by incorporating vorticity into our fluid model. Although the waves we study arise from a linearisation of the governing equations, the dynamical systems which prescribe the resulting fluid motion are themselves nonlinear.  We apply a phase-plane analysis in order to comprehensively describe qualitative properties of the underlying wave-field kinematics for various vorticity regimes.

\section{Equations of motion}
We consider here a two-dimensional water flow, moving under the influence of gravity, and use a reference frame with horizontal $x$ axis, and $y$ axis that points vertically upwards. The water flow occupies the domain bounded below by the rigid flat bed $y=-h$, (with $h>0$), and above by the free surface $y=\eta_1(x,t)+h_1$, the latter being a perturbation of the flat free
surface $y=h_1$. Here, $h_1>0$ is a constant, while $\eta_1(x,t)$ is a periodic function in $x$ of period $L>0$ which has mean zero, that is, for all $t,$
$$\int_{0}^L \eta_1(x,t) dx =0.$$
The fluid is assumed to be stratified into two layers of (differing) piecewise-constant densities, and so $\bm{\rho}=\rho_1$ on the upper fluid layer $\Omega_1(\eta,\eta_1)$, where
$$\Omega_1(\eta,\eta_1):=\{(x,y):x\in [0,L],\eta(x,t)<y<h_1+\eta_1(x,t)\},$$
while  $\bm{\rho}=\rho$ on the lower fluid layer $\Omega(\eta)$, where $$\Omega(\eta):=\{(x,y):x\in [0,L],-h<y<\eta(x,t)\}.$$
Here $y=\eta(x,t)$ denotes the internal interface separating the two fluid layers, which we also take to be periodic, $\eta(x+L,t)=\eta(x,t)$, and with zero mean,
\[  \int_0^L \eta(x,t) dx =0 \] for all $t.$ It is assumed that the fluid is stably stratified, and so $\rho_1< \rho\equiv (r+1)\rho_1$ for some $r>0$ (in an oceanographical context  $r=\mathcal O (10^{-3})$ would  typically be a reasonable value  \cite{CI}).  
With respect to the above stratification, the velocity field can be split into
\begin{equation}
\bm{u}(x,y,t):=
\left\{\begin{array}{ccc}  u(x,y,t), &{\rm in} & \Omega(\eta),\\ u_1(x,y,t), &{\rm in} &\Omega_1(\eta,\eta_1),\end{array}\right.
\end{equation}
and 
\begin{equation}
\bm{v}(x,y,t):=
\left\{\begin{array}{ccc} v(x,y,t), &{\rm in} & \Omega(\eta),\\ v_1(x,y,t), &{\rm in} &\Omega_1(\eta,\eta_1).\end{array}\right.
\end{equation}
\begin{figure}[H]
\begin{center}
 \resizebox{.8\textwidth}{!}{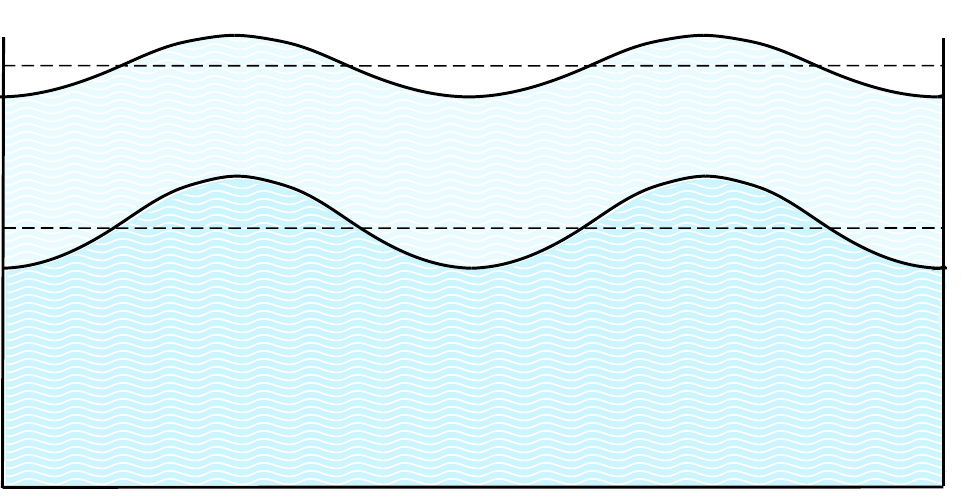}
\caption{Coupled surface-internal water waves for $a/a_1>0$, and $\gamma_1<0$.}
\label{Fig1}
\end{center}
\end{figure}
For ocean gravity waves it is reasonable to assume inviscid flow, whereby the equations of motion are given by the Euler equation.
Working in the two-dimensional setting, the governing equations can be expressed as
\begin{equation}\label{euler}
 \left\{\begin{array}{lcl}
        \bm{ u}_t+\bm{u}\bm{u}_x+\bm{v}\bm{u}_y &=& -\bm{\frac{1}{\rho}}P_x,\\
         \bm{v}_t+\bm{u}\bm{v}_x+\bm{v}\bm{v}_y  &=& -\bm{\frac{1}{\rho}}P_y-g,
        \end{array}\right.
\end{equation}
where $P=P(x,y,t)$ denotes the pressure, and $g$ is the gravitational acceleration.
It is also reasonable to assume incompressibility for ocean waves, and so
\begin{equation}\label{masscons}
\bm{ u}_x+\bm{v}_y=0\,\,\textrm{in}\,\,\Omega\cup\Omega_1.
\end{equation}
The vorticity of the fluid is defined to be $\bm \gamma=\nabla \times \bm u$, which in two-dimensions reduces  to 
\begin{equation}\label{vort2d}
\gamma_1=u_{1,y}-v_{1,x} \mbox{ in } \Omega_1, \mbox{ while } \gamma=u_{y}-v_{x}  \mbox{ in } \Omega.
\end{equation}
 In this paper we consider the physical regime whereby the upper fluid layer $\Omega_1(\eta,\eta_1)$ has constant (nonzero) vorticity, $\gamma_1\neq 0$, while the lower fluid layer  $\Omega(\eta)$ is irrotational, $\gamma=0$.
\begin{figure}[H]
\begin{center}
 \resizebox{.8\textwidth}{!}{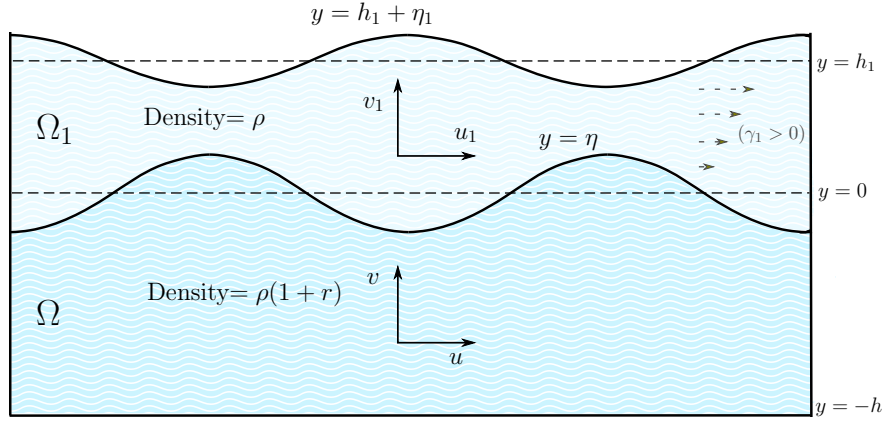}
\caption{Coupled surface-internal water waves for $a/a_1<0$, and $\gamma_1>0$.}
\label{Fig2}
\end{center}
\end{figure}
\noindent Complementing the equations of motion are the boundary conditions. The dynamic boundary condition 
\begin{equation}\label{atm}
 P=P_{atm}\,\,{\textrm on}\,\, y=\eta_1 (x)+h_1
\end{equation}
(with $P_{atm}$ being the constant atmospheric pressure) decouples the motion of the water from that of the air. The kinematic boundary conditions
 reflect the impermeability of the flat bed, the interface $y=\eta(x,t)$, and the free surface $y=h_1+\eta_1(x,t)$, and read as
\begin{equation}
 \label{kin_fs}
 v_1=\eta_{1,t}+u_1\eta_{1,x}\ \  {\rm on }\ \ y=\eta_1(x,t)+h_1,
\end{equation}
and
\begin{equation}
 \label{kin_int}
 \begin{array}{lll}
v_1=\eta_t +u_1 \eta_x & \ {\rm on } & y=\eta(x,t),\\
v=\eta_t+u\eta_x & {\rm on} & y=\eta(x,t),
\end{array}
\end{equation}
and 
\begin{equation}\label{dbc-b}
 v=0 \quad{\rm on}\quad y=-h.
\end{equation}
As a consequence of \eqref{masscons}, the (generalized) velocity potential $$\bm{\varphi}(x,y,t)=\left\{\begin{array}{ccc} \varphi(x,y,t) &{\rm in}& \Omega,\\ \varphi_1(x,y,t) &{\rm in} & \Omega_1,\end{array}\right.$$
can be defined by
\begin{equation}\label{vel_pot}
 \left\{\begin{array}{ll}
  u=\varphi_x,\quad  v=\varphi_y & {\rm in}\quad \Omega, \\
  u_1=\varphi_{1,x}+\gamma_1 y, \quad v_1=\varphi_{1,y} & {\rm in}\quad \Omega_1,
 \end{array}\right.
\end{equation}
where the potential functions $\varphi(x,t),$ $\varphi_1(x,t)$ are also $L$-periodic in $x$.
The kinematic boundary conditions \eqref{kin_fs} and \eqref{kin_int} can now be reexpressed as 
\begin{equation}\label{kin_surface}
 \eta_{1,t}=(\varphi_{1,y})_{s_1}-\eta_{1,x} [(\varphi_{1,x})_{s_1}+ \gamma_1 (h_1+\eta_1)]\end{equation}
and 
\begin{equation}\label{kin_interface}
   \eta_t=( \varphi_{1,y})_{s}-\eta_x [ (\varphi_{1,x})_{s} +\gamma_1 \eta], \qquad
 \eta_t=(\varphi_y)_{s}- \eta_x (\varphi_x)_{s},
\end{equation}
where the subscript $s_1$ denotes the trace of the involved functions on the free surface $y=\eta_1(x,t)+h_1$, while the subscript $s$ denotes the trace on the interface $y=\eta(x,t)$.
In the following we denote
\begin{align}\label{Phi_notation}\nonumber
 \Phi(x,t)&=\varphi(x,\eta(x,t),t),\qquad 
 \Phi_1(x,t)=\varphi_1(x,\eta(x,t),t),\\
 \Phi_2(x,t)&=\varphi_1(x,h_1+\eta_1(x,t),t).
\end{align}

\section{Linear approximation of the equations}
\subsection{Hamiltonian formulation}
The governing equations \eqref{euler}---\eqref{dbc-b} can be linearised by considering their Hamiltonian formulation, where the Hamiltonian functional for our physical set-up is given by the total energy
$$H=\int\int_{\Omega\cup\Omega_1}\bm{\rho}\left\{\frac{\bm{u}^2+\bm{v}^2}{2}+gy\right\}dydx.
$$
 The Hamiltonian, $H$, is given by the energy of the system for one period (that is, over horizontal length $L$). The Hamiltonian density of the Hamiltonian functional can be written explicitly by following an approach based on Dirichlet-Neumann (DN) operators, see \cite{CGK}. 
A detailed account of the Hamiltonian formulation  for the more complicated setting of constant (non-zero)  vorticity in each layer is available in \cite{CI,CIM} (however note that an integral density is used in these works since solutions are considered over the entire real-line).

In order to obtain a systematic linearisation of the governing equations we perform a suitable nondimensionalisation and scaling of the variables. To simplify the derivation process, we make the assumption that the depths $h$ and $h_1$ are roughly of the same order of magnitude, so we can introduce a depth scale $ \mathfrak {h}$ such that $h,h_1= \mathcal O(\mathfrak{h})$; similarly for the  amplitudes of the surface and internal wave, $a$ and $a_1$, respectively --- we introduce the amplitude length scale $ \mathfrak {a}$ and assume that  $a,a_1= \mathcal O(\mathfrak{a})$. Defining the dimensionless parameters
\begin{equation}\label{param}
\epsilon= \mathfrak {a}/ \mathfrak {h}, \quad  \delta= k\mathfrak {h}, \quad  \mathfrak e=\mathfrak ak
\end{equation}
where $k=2\pi/L>0$ is the wavenumber, it follows that $\epsilon$  measures the amplitude size relative to the layer depth, $\delta$ measures the depth relative to the wavelength, while $\mathfrak e$ is the {\em wave steepness} parameter. In the linear regime it is necessary for both $\epsilon$ and $\mathfrak e$ to be very small ($\ll1$), while for the time being we can assume that $\delta = \mathcal{O}(1)$.
These parameters can be used to nondimensionalise all quantities along classical lines (cf. \cite{Johnson})
\begin{align}
\bar{ h} &=  \mathfrak{h} h ,  \quad  \bar{ h}_1 =  \mathfrak{h} h_1 ,  \quad \bar{x}= \frac{\mathfrak{h}}{\delta} x, \quad \bar{y}= \frac{\mathfrak{h}}{\delta^2} y ,\quad \bar{t} = \frac{\mathfrak{h} }{\delta \sqrt{\bar{g}\mathfrak{h}}}  t, \quad \bar{\eta} = \epsilon \mathfrak{h} \eta,  \nonumber \\
\bar{\varphi}&= \frac{\epsilon}{\delta } \mathfrak{h } \sqrt{\bar{g}\mathfrak{h}} \varphi, \quad \bar{\gamma}_1 =  \frac{ \sqrt{\bar{g}\mathfrak{h}} }{\mathfrak{h}}   \gamma_1, \quad 
\bar{\eta}_1 = \epsilon \mathfrak{h} \eta_1, \quad  \bar{\varphi}_1= \frac{\epsilon}{\delta } \mathfrak{h } \sqrt{\bar{g} \mathfrak{h}} \varphi_1.
\end{align}
Here the dimensional quantities are those with a bar, while the nondimensional ones are without. The nondimensional version of the Earth's acceleration $\bar{g}=9.81$ m/s$^2$ is  $g=1$, but we will retain $g$ throughout the following in order to easily revert back to the dimensional form of the governing equations. Then equations 
\eqref{kin_surface}--\eqref{Phi_notation} become (for the non-dimensional variables)
\begin{equation}\label{kin_surface-0}
 \eta_{1,t}=(\varphi_{1,y})_{s_1}-\eta_{1,x} [\epsilon (\varphi_{1,x})_{s_1}+ \gamma_1 (h_1+ \epsilon \eta_1)]\end{equation}
\begin{equation}\label{kin_interface-0}
   \eta_t=( \varphi_{1,y})_{s}-  \epsilon \eta_x [ (\varphi_{1,x})_{s} +\gamma_1 \eta],\
 \eta_t=(\varphi_y)_{s}- \epsilon \eta_x (\varphi_x)_{s}   ,
\end{equation}
\begin{align}\nonumber
 \Phi(x,t)&=\varphi(x,\epsilon \eta(x,t),t),\quad 
 \Phi_1(x,t)=\varphi_1(x, \epsilon \eta(x,t),t),\\ \label{Phi_notation-0}
 \Phi_2(x,t)&=\varphi_1(x,h_1+ \epsilon \eta_1(x,t),t).
 \end{align} 
In the leading order, as $\epsilon \to 0$, equations
\eqref{kin_surface-0}--\eqref{Phi_notation-0} are approximated  by
\begin{equation}\label{kin_surface1}
 \eta_{1,t}=(\varphi_{1,y})_{s_1} -   \gamma_1 h_1 \eta_{1,x},
 \end{equation}
\begin{equation}\label{kin_interface1}
   \eta_t=( \varphi_{1,y})_{s}, \quad 
 \eta_t=(\varphi_y)_{s},
\end{equation}
and
\begin{align}\label{Phi_notation1}
 \Phi(x,t)=\varphi(x,0,t),\quad
 \Phi_1(x,t)=\varphi_1(x,0,t),\quad
 \Phi_2(x,t)=\varphi_1(x,h_1,t).\end{align}
The resulting Hamiltonian variables are $\eta, \eta_1$ and
\begin{equation} \label{HV1}
\begin{split}
&\xi=\frac{\rho \varphi(x,0,t) -\rho_1 \varphi_1(x,0,t)}{\rho_1}\equiv (r+1)\Phi(x,t)-\Phi_1(x,t)  , \\
& \xi_1= \varphi_1(x,h_1,t)   \equiv   \Phi_2(x,t).
\end{split}
\end{equation}
  With the nondimensionalisation and scaling introduced above, and ignoring the linear terms (whose average is $0$), the Hamiltonian admits an expansion of the form
\begin{equation} \label{Hexp}
\frac{\bar{H}}{ \bar{\rho}_1 \bar{g} \mathfrak{h}^3}= H(\xi,\xi_1,\eta,\eta_1)=\epsilon^2 H^{(2)}(\xi,\xi_1,\eta,\eta_1) +\epsilon^3 H^{(3)}(\xi,\xi_1,\eta,\eta_1)+ \ldots.
\end{equation}
The leading order term is (see formula  (4.15) in  \cite{CI} with $\gamma=0$)
\begin{align*}
&H^{(2)}(\xi,\xi_1, \eta, \eta_1)=\frac{1}{2}\int_{0}^L\Big[ \xi \frac{D\tanh(hD)\coth(h_1D)}{(r+1) \coth(h_1D)+  \tanh(hD)}\xi \\
&\ +2\xi \frac{D\tanh(hD)\text{csch}(h_1D)}{(r+1) \coth(h_1D)+  \tanh(hD)}\xi_1+\xi_1 \frac{D [\tanh(hD)\coth(h_1D)+r+1 ]}{(r+1) \coth(h_1D)+  \tanh(hD)}\xi_1 \\
&\qquad\qquad\qquad\qquad\qquad\qquad\qquad\qquad + g r \eta^2+  ( g+ \gamma_1 ^2 h_1) \eta_1^2 - 2 \gamma_1 h_1 \xi_1 \eta_{1,x} \Big]dx,
\end{align*}
where the operator $D$ is defined as  $D=-i\partial_x$.
The corresponding Hamiltonian equations have the form (cf.  \cite{CI, CIM})
\begin{equation}
\begin{split}
\epsilon^2 \xi_t=&-\frac{\delta H}{\delta \eta}-\Gamma\int_{0}^{x}\frac{\delta H}{\delta \xi(x')}dx', \quad \qquad 
\epsilon^2 \eta_t=\frac{\delta H}{\delta \xi}\\
\epsilon^2 \xi_{1,t}=&-\frac{\delta H}{\delta \eta_1}-\Gamma_1\int_{0}^{x}\frac{\delta H}{\delta \xi_1(x')}dx', \quad \quad
\epsilon^2 \eta_{1,t}=\frac{\delta H}{\delta \xi_1}
\end{split}
\end{equation}
where $\Gamma=  -  \gamma_1$, $\Gamma_1= \gamma_1 = - \Gamma$
are constants. At the leading order, $H= \epsilon^2 H^{(2)}$, these equations are linear, and integration is replaced by $\partial_x^{-1}$, giving
\begin{align}
\nonumber
\xi_t=&-g r \eta-\Gamma \partial_x^{-1}\eta_t, \\
\nonumber
 \eta_t=& \frac{D\tanh(hD)\coth(h_1D)}{(r+1) \coth(h_1D)+  \tanh(hD)}\xi 
 \\
 & \label{linsys11} \qquad \qquad \qquad \qquad \qquad + \frac{D\tanh(hD)\text{csch}(h_1D)}{(r+1) \coth(h_1D)+ \tanh(hD)}\xi_1  , \\
\nonumber
\xi_{1,t}=&- \gamma_1 h_1 \xi_{1,x}-  (g+\gamma_1^2 h_1)\eta_1 -\Gamma_1 \partial_x^{-1}\eta_{1,t} ,   \\
\nonumber
\eta_{1,t}=&-\gamma_1 h_1 \eta_{1,x}+\frac{D\tanh(hD)\text{csch}(h_1D)}{(r+1) \coth(h_1D)+ \tanh(hD)}\xi
\\ 
\nonumber & \qquad  \qquad \qquad \qquad \qquad  + \frac{D [\tanh(hD)\coth(h_1D)+r+1]}{(r+1) \coth(h_1D)+ \tanh(hD)}\xi_1 .
\end{align}
These equations will be used later to relate the boundary values of the potentials on the surface and the internal interface and to obtain matching conditions, which will lead to a definition \eqref{A} of the parameter $A$, and to the dispersion relation for our system.
\subsection{Wave solutions}
The aim now is to explicitly determine the velocity potentials in the body of the fluid from  their values on the surface and on the interface. Essentially, this procedure is as an analytic continuation based on the fact that the potentials satisfy the Laplace equation inside the fluid, which follows from \eqref{masscons}, \eqref{vort2d} and \eqref{vel_pot}. The boundary values will be obtained from the linear equations (in the leading order) that relate them to $\eta_t$ and $\eta_{1,t}$, namely \eqref{kin_surface1} and \eqref{kin_interface1}. The relevant boundary value problem for the lower layer is
\begin{equation} \label{LL1}
\begin{split}
&\Delta \varphi(x,y) =0, \\
&\frac{\partial \varphi}{\partial \mathbf{n}} =0 \,\, \text{on} \,\, y=-h,\\
& \varphi_y(x,0,t)=\eta_t(x,t),
 \end{split}
\end{equation}
where $\mathbf{n}$ denotes the outward pointing unit normal, and differentiation represents the normal derivative. 
In our considerations it must be taken into account that, in the periodic case, the set of eigenvalues of the operator $D=-i\partial_x$ is discrete. Indeed, the operator $D$ now has eigenvalues $\tilde{k}=\tilde{k}_n=\frac{2\pi n}{L}$ for any integer $n,$ and periodic eigenfunctions $e^{i\tilde{k}x},$ since
\[  
D e^{i \tilde{k}x} = \tilde{k} e^{i\tilde{k}x}. 
\]
For convenience,  summation over the integers $n$ (from $-\infty$ to $\infty$) will be written  as summation over $\tilde{k}$.
The dependence on $t$ is not written explicitly (we can treat $t$ here as a parameter) and the problem is stated in the $(x,y)$ domain. 
The general solution in the bulk is a superposition of harmonic functions $\varphi_{\tilde{k}}(x,y),$  periodic in $x,$ satisfying \eqref{LL1}:
\begin{equation} \label{LL2}
    \varphi(x,y)=\sum_{\tilde{k}} \mathcal{A}(\tilde{k}) \varphi_{\tilde{k}}(x,y) , \quad \varphi_{\tilde{k}}(x,y)=e^{i \tilde{k}x}\cosh [\tilde{k}(y+h)].
\end{equation}
The as yet unknown function $\mathcal{A}(\tilde{k})$ can be obtained from the interface condition
\begin{equation} \label{LL3}
\eta_t(x)=\varphi_y(x,0)=\sum_{\tilde{k}}  \tilde{k}\mathcal{A}(\tilde{k}) e^{i\tilde{k}x}\sinh (\tilde{k} h) .
\end{equation}
The Fourier coefficient of the above expansion is given by
\begin{equation} \label{LL4}
\begin{split}
&\mathcal{A}(\tilde{k})=\frac{1}{L \tilde{k}\sinh (\tilde{k} h) }\int_0^L e^{-i\tilde{k}x'}\eta_t(x')dx',\\
&\varphi(x,y)=\frac{1}{L}\sum_{\tilde{k}}\int_0 ^L e^{i\tilde{k}(x-x')} \eta_t(x') \frac{\cosh[\tilde{k}(y+h)]}{\tilde{k}\sinh(\tilde{k}h)} dx'.
\end{split}
\end{equation}
The periodic delta-function 
$$\delta(x-x')=\frac{1}{L}\sum_{\tilde{k}} e^{i\tilde{k}(x-x')}  $$ 
is normalised by the condition $\int_0^L \delta(x) dx=1$ and we have
\begin{equation} \label{LL5}
\begin{split}
&\varphi(x,y)= \frac{\cosh[D(y+h)]}{D\sinh(Dh)} \int_0 ^L \frac{1}{L} \sum_{\tilde{k}} e^{i\tilde{k}(x-x')} \eta_t(x') dx' = \frac{\cosh[D(y+h)]}{D\sinh(Dh)} \eta_t(x).
\end{split}
\end{equation}
The Ansatz which is typically invoked for linear water waves is
\begin{equation} \label{HV2}
\eta=a\cos(kx-\omega t), \qquad \eta_1= a_1\cos(kx-\omega t),
\end{equation}
and, for such sinusoidal $\eta$ and $\eta_1$, we note that the action of `$D$' may be formally replaced by multiplication with `$k$' for operators which involve only even orders of $D$. Therefore 
\begin{equation}\label{phisol}
    \varphi(x,y,t)= a c(k) \frac{\cosh[k(y+h)]}{\sinh(kh)} \sin(kx-\omega t),
\end{equation}
and also 
\begin{equation} \label{Phi}
    \Phi(x,t):=\varphi(x,0,t)=ac \coth(kh) \sin(kx-\omega t),
\end{equation}
where \begin{equation}
    c(k)= \frac{\omega(k)}{k}
\end{equation} is the phasespeed of the wave.
The boundary-value problem in the upper layer is
\begin{equation} \label{UL1}
\begin{split}
&\Delta \varphi_1(x,y) =0, \\
&\varphi_{1,y} (x, h_1)=\eta_{1,t}(x)+\gamma_1 h_1 \eta_{1,x} \,\, \text{on} \,\, y=h_1,\\
& \varphi_{1,y}(x,0)=\eta_t(x).
 \end{split}
\end{equation}
The general solution in the bulk is a superposition of harmonic functions of the form $e^{\pm \tilde{k}y+i\tilde{k}x},$ periodic in $x,$ that is
\begin{equation} \label{UL2}
\varphi_1(x,y)=\sum_{\tilde{k}} \left(\mathcal{A}(\tilde{k})e^{\tilde{k}y+i\tilde{k}x}+\mathcal{B}(\tilde{k})e^{-\tilde{k}y+i\tilde{k}x} \right).
\end{equation}
The as yet unknown Fourier coefficients $\mathcal{A}(\tilde{k})$ and $\mathcal{B}(\tilde{k})$ can be obtained from the surface and interface conditions
\begin{equation} \label{UL3}
\begin{split}
&\eta_{1,t}+ \gamma_1 h_1 \eta_{1,x}=\varphi_{1,y}(x,h_1)=\sum_{\tilde{k}} e^{i\tilde{k}x} \tilde{k} \left( \mathcal{A}(\tilde{k})e^{\tilde{k}h_1} -\mathcal{B}(\tilde{k})e^{-\tilde{k}h_1} \right),\\
&\eta_t=\varphi_{1,y}(x,0)=\sum_{\tilde{k}} e^{i\tilde{k}x}  \tilde{k} \left( \mathcal{A}(\tilde{k}) -\mathcal{B}(\tilde{k})\right).
\end{split}
\end{equation}
Evaluating the coefficients of the Fourier series given above allows us to determine $\mathcal{A}(\tilde{k})$ and $\mathcal{B}(\tilde{k}):$
\begin{equation} \label{UL4}
\begin{split}
&\mathcal{A}(\tilde{k})= \frac{1}{L}\int_{0}^L e^{-i\tilde{k}x'}\frac{[\eta_{1,t}(x')+ \gamma_1 h_1 \eta_{1,x}(x')]-e^{-\tilde{k}h_1}\eta_t(x')}{2\tilde{k}\sinh(\tilde{k}h_1)} dx',\\
& \mathcal{B}(\tilde{k})= \frac{1}{L}\int_0^L e^{-i\tilde{k}x'}\, \frac{[\eta_{1,t}(x')+ \gamma_1 h_1 \eta_{1,x}(x')]-e^{\tilde{k}h_1}\eta_t(x')}{2\tilde{k}\sinh(\tilde{k}h_1)} dx',
\end{split}
\end{equation}
and proceeding as in the solution of the previous problem we obtain finally
\begin{equation} \label{UL5}
\varphi_1(x,y)=\frac{\cosh(yD)}{D\sinh (h_1D)} [\eta_{1,t}(x)+ \gamma_1 h_1 \eta_{1,x}(x)]-\frac{\cosh((y-h_1)D)}{D\sinh( h_1D)}\eta_t(x),
\end{equation}
which the Ansatz \eqref{HV2} gives
\begin{align}
\varphi_1(x,y,t)=a_1\cdef\left\{\sinh k(y-h_1)+A\cosh k(y-h_1) \right\} \sin (kx-\omega t) \nonumber
\\ \mbox{ in } 0<y<h_1, \label{phi1sol} 
\end{align}
for 
\begin{equation}\label{modc}
\cdef=c-\gamma_1 h_1
\end{equation} 
and 
\begin{equation}\label{A}
A:=-\frac {a}{a_1}\frac{c}{\cdef \sinh kh_1}+\frac{\cosh kh_1}{\sinh kh_1}.
\end{equation}
\begin{remark}
The quantity $\cdef$ measures the speed of the sheared current at the surface of the upper layer relative to the phasespeed of the wave.  The sign of  $c\cdot \cdef$ is linked to the appearance of critical layers. The Rayleigh equation for linear fluid flow in the upper-layer is
\[
(\gamma_1 y -c ) \left(-\tilde v''(y) + k^2 \tilde v(y)\right)=0,
\]
and this equation becomes singular if  $\gamma_1 y^{crit}=c$ for some value $y^{crit}\in (0,h_1)$, near which depth critical layers occur. 
Thus, critical layers occur in the upper fluid layer if and only if $c\cdot\cdef<0$. In the setting of a single fluid layer, it was first observed by Kelvin that streamlines resembling cat's eye patterns form  near the critical layer; we show in Section \ref{sec-PP} below that a similar phenomenon occurs also in the two-fluid layer setting near such critical values.
\end{remark}
On the boundaries of the upper domain we have
\begin{equation}\label{Phi1}
\begin{split}
    &\Phi_1(x,t):= \varphi_1(x,0,t)=a_1\cdef\left [-\sinh (k h_1)+A\cosh (k h_1) \right] \sin (kx-\omega t),\\
    &\Phi_2(x,t):= \varphi_1(x,h_1,t)=a_1\cdef A \sin(kx-\omega t),
\end{split}
    \end{equation}
and substituting \eqref{HV2} into  the first and the third equations of \eqref{linsys11} gives
\begin{equation} \label{HV3}
 \xi =\frac{a}{k c } ( g r + \gamma_1 c ) \sin(kx-\omega t), \quad  \xi_1 =\frac{ a_1 [g- \gamma_1 \cdef]}{k \cdef}  \sin(kx-\omega t).
\end{equation}
Recalling the definition of Hamiltonian variables \eqref{HV1}, \eqref{Phi} and \eqref{Phi1} gives
\begin{align*}
  \xi= [(r+1) a c \coth(k h) -  a_1 \cdef(-\sinh(k h_1)+A\cosh(k h_1))]  \sin(kx-\omega t),
\end{align*} and  comparison with \eqref{HV3} leads to
\begin{align}\label{C1}
  [(r+1) a c \coth(k h) -  a_1\cdef(-\sinh(k h_1)+A\cosh(k h_1))]=  \frac{a}{k c }\left[ g r +  \gamma_1 c \right ] .
\end{align}
From \eqref{HV1}, \eqref{Phi1} and \eqref{HV3} we obtain an expression for $A$ with no explicit dependence on $a/a_1$:
 \begin{equation} \label{AA}
    A=\frac{g-\gamma_1\cdef}{k\cdef^2}.
\end{equation}
{\begin{remark}\label{A-pos-rem}
The  nondimensional parameter $A$ plays a key role in determining the qualitative features of the resulting fluid motion. In  \cite{HV-JDE,HV-AN} it was shown that in the irrotational setting ($\gamma_1=0$) the parameter ranges $0<A<1$, $A=1$, and $A>1$  prescribe regimes with  qualitatively different underlying flow kinematics. However, recently in \cite{HIV}, it was proven that $A$ can, in fact, never be less than $1$ in the irrotational setting. In the rotational setting considered herein, $A$ can take values in the ranges $A>1$, $A=1$, and $0<A<1$, and indeed it follows from \eqref{AA} that $A$ may also take negative values. Discussion of the asymptotic behaviour of $A$  is given in the Appendix.
\end{remark}}
\begin{lemma}\label{ALemma}
\begin{itemize}
\item[(i)] For $\gamma_1\leq 0$, then $A>0$ is strictly positive.  
\item[(ii)] For $\gamma_1>0$ and $c<\gamma_1h_1$ ($\cdef<0$),  then $A>0$.  
\item[(iii)] For $\gamma_1>0$, $c>\gamma_1h_1$ ($\cdef>0$), and $c^2< 4gh_1$, then $A>0$.  
\item[(iv)] For $\gamma_1>0$, $c>\gamma_1h_1$ ($\cdef>0$),  and $4gh_1\leq c^2$,  then $A\leq 0$ is non-positive.  
 \end{itemize}
 \end{lemma}
 \begin{proof}
The proof of the first two parts follows immediately from \eqref{AA}, since $k>0$. For part $(iii)$  we note that \eqref{AA} implies that $A$ has the same sign as $\gamma_1^2h_1-\gamma_1 c+g$, and this is strictly positive if $c^2< 4gh_1$. Part $(iv)$ follows by noting that this quadratic in $\gamma_1$ is negative if $c^2> 4gh_1$, since we assume that $c>\gamma_1h_1$. (Note that $4g<\gamma_1 c$ implies that $c^2>4gh_1$, for $c>\gamma_1h_1$.) Relation \eqref{AA}, regarded as a quadratic in $\cdef$,  has real-valued solutions if and only if $A\geq -\frac{\gamma_1^2}{4gk}$, which serves as a  lower bound for the parameter $A$.
 \end{proof}
It is useful for later reference to reexpress \eqref{A} as
\begin{equation}\label{A-alt}
\frac {a}{a_1}\frac{c}{\cdef}=\cosh kh_1 -A \sinh kh_1=\frac{1-A}{2}e^{kh_1}+\frac{1+A}{2}e^{-kh_1}
\end{equation}
where the right-hand side is a strictly decreasing function of $A$. 
From  \eqref{AA} and \eqref{A-alt} we obtain
\begin{equation} \label{RR}
\begin{split}
 \frac{a}{a_1}= \frac{\cdef}{c}\sinh(k h_1)\left(\coth(k h_1) - \frac{g-\gamma_1\cdef }{k\cdef^2} \right).
\end{split}
\end{equation}
Regarding \eqref{RR} as a quadratic in $\cdef$ then, if $\frac{a}{a_1}\frac{c}{\cdef}>0$,  
\[
-\frac{\gamma_1 \tanh kh_1}{2k}+\frac{\sqrt{\gamma_1^2 \tanh^2 kh_1 + 4gk \tanh kh_1}}{2k}<\cdef,
\]
for $c>\gamma_1h_1$ $(\cdef>0)$, while
\[
\cdef<-\frac{\gamma_1 \tanh kh_1}{2k}-\frac{\sqrt{\gamma_1^2 \tanh^2 kh_1 + 4gk \tanh kh_1}}{2k},
\]
for $c<\gamma_1h_1$ $(\cdef<0)$. When $\frac{a}{a_1}\frac{c}{\cdef}<0$, there is a reversal of sign in the above inequalities. In the presence of a critical layer ($c/\cdef<0$), if ${a}/{a_1}<0$ then
\[
\gamma_1>\sqrt{\frac{g \tanh kh_1}{kh_1^2 -h_1 \tanh kh_1}},
\]
as can be seen from the second inequality for $\cdef$ above. It is interesting to note that this expression matches the necessary and sufficient condition that ensures the existence of critical layers in the one fluid layer setting, cf \cite{CSV,DP}.
From \eqref{C1} and \eqref{AA} we have
\begin{align}\label{C2}
    a& \left( (r+1) c \coth (k h) - \frac{g r}{kc} - \frac{\gamma_1}{k} \right) \nonumber \\
    &\qquad \qquad \qquad = a_1\cdef\left(-\sinh(k h_1)+\frac{g-\gamma_1\cdef }{k\cdef^2}\cosh(k h_1)  \right). 
\end{align}
\subsection{Dispersion relation}
Relations \eqref{RR} and \eqref{C2} can be used to obtain two different expressions for the ratio of amplitude parameters $a/a_1$: equating these expressions results in the dispersion relation
\begin{align}
\nonumber & P_4(c) := \left(1+\frac{1+r}{\tanh{kh}\tanh kh_1}\right) c^4 
-\gamma_1h_1\left[2+\frac{1+r}{\tanh{kh}}\left(\frac 2 {\tanh{kh_1}}-\frac 1 {kh_1}\right)\right] c^3
\end{align}
\begin{align}
& -\left[\frac {g  (1+r)} k \left(\frac 1 {\tanh{k h_1}} + \frac 1 {\tanh{kh}}\right)\right.\nonumber
\\& \nonumber
\qquad \qquad \left.+(\gamma_1 h_1)^2\left\{\left(\frac 1 {kh_1} + \frac {1+r} {\tanh{kh}}\right)\left(\frac 1 {kh_1} - \frac 1 {\tanh{kh_1}}\right)-1\right\}\right] c^2 
\end{align}
\begin{align} &-\frac{\gamma_1 h_1} k\left[\gamma_1^2 h_1 \left(\frac 1 {\tanh{k h_1}} - \frac 1 {kh_1}\right) + g\left(\frac {r-1} {kh_1} - \frac {2r} {\tanh{kh_1}}\right) \right] c \nonumber
\\  & \qquad \qquad \qquad \qquad  -r \frac g k \left[(\gamma_1 h_1)^2\left( \frac  1 {\tanh{kh_1}} - \frac 1 {kh_1}\right) - \frac g k\right]=0. \label{PressInt2}
\end{align}
A detailed discussion of the roots of this equation, and the corresponding wave solutions, in the context of equatorial wave-current interactions  can be found in  \cite{CI}.
 \begin{remark}
When $r\to 0$, the two layers have the same density. Disregarding the zero solution, $c=0$, the dispersion relation will degenerate to a cubic equation, namely
\begin{align*}
&\left[-k c( \coth kh_1+ \coth kh )+ \gamma_1  \right]\left( k\cdef^2 \coth kh_1 +  \gamma_1 \cdef - g\right) \qquad \qquad \qquad 
\\&
\qquad \qquad \qquad \qquad \qquad  \qquad \qquad \qquad  \qquad   \qquad + \frac{1}{\sinh^2 kh_1}k^2\cdef^2 \, c=0. 
\end{align*}
This matches  the cubic dispersion relation  in \cite{MM}, which is also posed in terms of $\cdef$, following various rearrangements of terms. 
\end{remark}
\begin{remark}
    In the limit where the lower layer is infinitely more dense than the upper one, $r \to \infty$, the dispersion relation is reduced to:
\[
\left( -k c^2 \coth kh  +g \right)\left( k\cdef^2 \coth kh_1 +  \gamma_1 \cdef - g\right)=0,
\]
with solutions $c_1^2=\frac{g\tanh kh}{k}$ and $\cdef_{1}^{\pm}=-\frac{\gamma_1 \tanh kh_1}{2k}\pm \frac{\sqrt{\gamma_1^2 \tanh^2 kh_1 + 4 gk \tanh kh_1}}{2k}$. 
\end{remark}
\begin{remark}
Taking the limit $a\to 0$, the upper-layer now has just one nontrivial free-boundary. In particular, $A=\coth kh_1$ and \eqref{AA}, \eqref{C2} imply that either $\cdef=0$ ($c=\gamma_1 h_1$) or else
\[
c=\frac 1 {2k}\left[ \gamma_1 \left(2kh_1 - \tanh kh_1\right)\pm \sqrt{\gamma_1^2 \tanh^2 kh_1 + 4kg \tanh kh_1} \right].
\]
This matches the dispersion relation for waves propagating on a single homogeneous fluid layer, in the presence of a sheared background current \cite{CS-CPAM}.
\end{remark}
\begin{remark}\label{inv-prop}
The dispersion relation \eqref{PressInt2} is invariant under the transformation $(c,\gamma_1)\mapsto(-c,-\gamma_1)$.
\end{remark}
When  the constant term of the quartic equation \eqref{PressInt2} representing the dispersion relation is negative, we have at least  two real roots, in view of the positivity of the leading order coefficient. Thus, for $\gamma_1, h_1, k$ such that
\[
\left[(\gamma_1 h_1)^2\left( \frac 
 1 {\tanh{kh_1}} - \frac 1 {kh_1}\right) - \frac g k\right]>0\iff\frac{\tanh kh_1}{kh_1}<\frac{\gamma_1^2 h_1} {g + \gamma_1^2 h_1},
\]
we have at least two real solutions. Returning to \eqref{RR}, we obtain two amplitude ratios as well, provided that $\cdef\neq 0.$ From Vieta's formulas \cite{PP}, the product of all the roots of \eqref{PressInt2} is proportional to 
{\[
-r \frac g k  \left(\frac {\tanh{kh}\tanh kh_1}  {\tanh{kh}\tanh kh_1+1+r}\right) \left[\gamma_1^2 h_1\left( \frac 
 {kh_1 - \tanh kh_1} {k \tanh kh_1} \right) - \frac g k\right].
\]}
\begin{remark}
    In the long wave regime, we can assume that  the coefficient  of the quadratic term in \eqref{PressInt2} is negative, that is, that the following inequality holds true:
 \begin{multline*}
     \frac {g  (1+r)} k \left(\frac 1 {\tanh{k h_1}} + \frac 1 {\tanh{kh}}\right)\\
+ (\gamma_1 h_1)^2\left\{\left(\frac 1 {kh_1} + \frac {1+r} {\tanh{kh}}\right)\left(\frac 1 {kh_1} - \frac 1 {\tanh{kh_1}}\right)-1\right\}>0.
 \end{multline*}
Indeed, under the long wave assumption, we have $\tanh kh_1 \approx kh_1$, whereas $ \frac{g}{k}\gg 1$. In that setting, the quadratic expression
\begin{align*}
& P_4''(c) = 12 \left(1+\frac{1+r}{\tanh{kh}\tanh kh_1}\right) c^2 
\\ &- 6 \gamma_1h_1\left[2+\frac{1+r}{\tanh{kh}}\left(\frac 2 {\tanh{kh_1}}-\frac 1 {kh_1}\right)\right] c - 2 \left[\frac {g  (1+r)} k \left(\frac 1 {\tanh{k h_1}} + \frac 1 {\tanh{kh}}\right)\right.\nonumber\\&\left.+(\gamma_1 h_1)^2\left\{\left(\frac 1 {kh_1} + \frac {1+r} {\tanh{kh}}\right)\left(\frac 1 {kh_1} - \frac 1 {\tanh{kh_1}}\right)-1\right\}\right] 
\end{align*}
will have two distinct real roots, which we denote by $c_1'',c_2''$, that can be written down explicitly. If the condition $P_4'(c_1) P_4'(c_2)>0$ holds true, then the equation $P_4'(c)=0$ has at most one real solution. Thus, we can infer whether  \eqref{PressInt2} has simple real solutions. On the other hand, a condition like $\frac{\tanh kh_1}{kh_1}<\frac{\gamma_1^2 h_1} {g + \gamma_1^2 h_1}$ can only hold for $ \gamma_1\gg g/h_1$ in the long wave setting.
\end{remark}
 \begin{remark}
  For $\gamma_1=0$, the dispersion relation \eqref{PressInt2}  reduces to that of a system of  coupled surface and internal waves above a flat bed:
\begin{align*}
&\left[1+\frac{1+r}{\tanh{kh}\tanh kh_1}\right] c^4  -\left[\frac {g  (1+r)} k \left( \frac 1 {\tanh{k h_1}} + \frac 1 {\tanh{kh}}\right)\right] c^2 \nonumber +r \frac {g^2} {k^2} =0, \nonumber
\end{align*}
see \cite{HV-JDE}.
 \end{remark}
 \begin{remark}
 When $k\ll 1$, we can approximate $\tanh kh \approx kh$, $\tanh kh_1 \approx kh_1$, so that
 \begin{align*}
 c^4 &-\gamma_1h_1c^3- g  (h+  h_1) c^2 +  \gamma_1 h h_1 g  c + \frac{r g^2 h h_1}{ 1+r } \approx 0 \nonumber,
\end{align*}
a relation also derived in \cite{CI}.
 \end{remark}
The remainder of this paper is  concerned with determining qualitative properties of the fluid motion which is prescribed by the above wave solutions, by subjecting the relevant dynamical systems to a phase plane analysis. We do not address matters concerning the generation, or stability (see, for instance, \cite{BV}), of such wave solutions.


\section{Phase plane analysis}\label{sec-PP}
 Explicit expressions for the velocity fields induced by the linear wave Ansatz \eqref{HV2}  in the lower, and upper, fluid layers can be obtained directly from  \eqref{vel_pot},  \eqref{phisol} and \eqref{phi1sol}. If $(x(t),y(t))$ is the path of a particle in the lower-fluid layer $\Omega$, then the motion of the particle is described by the nonlinear nonautonomous dynamical system 
\begin{subequations}\label{DYN-1}
\begin{align}[left ={\empheqlbrace}]
\nonumber
 \frac{dx}{dt}&=u=a\omega\cos (kx-\omega t)\frac{\cosh k(y+h)}{\sinh kh} \\ \label{LowSys1}
 \frac{dy}{dt}&=v=a\omega \sin (kx-\omega t)\frac{\sinh k(y+h)}{\sinh kh},
\end{align}
for $-h<y<0$, with initial data $(x_0,y_0)$. In the upper-fluid layer $\Omega_1$,  particle trajectories $(x(t),y(t))$ are prescribed by the nonlinear nonautonomous dynamical system 
\begin{align}[left ={\empheqlbrace}]
\nonumber
 \frac{dx}{dt}&=a_1k\cdef \cos (kx-\omega t)\left\{\sinh k(y-h_1)+A\cosh k(y-h_1) \right\}   +\gamma_  1 y   \\ 
 \frac{dy}{dt}&=a_1k\cdef \sin (kx-\omega t)\left\{\cosh k(y-h_1)+A\sinh k(y-h_1) \right\},  \label{Upper-DS0}
\end{align}
\end{subequations}
for $0<y<h_1$, with initial data $(x_0,y_0)$. The mean-level of the oscillating internal wave interface $y=\eta$ is located at $y=0$, whereas the free-surface $y=h_1+\eta_1$ oscillates about the mean-level located at $y=h_1$. 
The right-hand sides of the differential systems \eqref{LowSys1} and \eqref{Upper-DS0}  are smooth, therefore the existence of unique local smooth solutions is ensured by the Picard--Lindel\"of theorem \cite{Meiss}. Furthermore, since $y$ is bounded, the right-hand sides of \eqref{LowSys1} and \eqref{Upper-DS0} are bounded, hence these unique solutions are defined globally \cite{Meiss}. The right-hand sides of both \eqref{LowSys1} and \eqref{Upper-DS0} are nonlinear, and thus such systems cannot be solved explicitly. Rather than linearising  the dynamical systems \eqref{LowSys1} and \eqref{Upper-DS0}, we will analyse the fully nonlinear dynamical systems by way of phase plane methods, thereby establishing the salient qualitative features of the flow prescribed by both  \eqref{LowSys1} and \eqref{Upper-DS0}. Since the fluid layers are separated by an impermeable interface $y=\eta(x,t)$, and the solutions \eqref{phisol} and \eqref{phi1sol} satisfy the matching conditions \eqref{kin_interface1} at this interface by design, we can address the phase plane analysis of system \eqref{LowSys1}  in the lower-fluid layer $\Omega$, and system \eqref{Upper-DS0} in the upper-fluid layer $\Omega_1$, separately in the first instance, and then piece together the information to get a picture of the motion of the entire two-layer body.

For the purposes of performing a phase-plane analysis of the nonlinear dynamical systems \eqref{LowSys1} and \eqref{Upper-DS0}, without loss of generality we can restrict our attention to waves for which $c>0$, noting from Remark \eqref{inv-prop} that if $c$ is a wavespeed prescribed by the dispersion relation \eqref{PressInt2} for $\gamma_1$, then $-c$ will be a solution of \eqref{PressInt2} for the vorticity $-\gamma_1$.
 We shall regard the wavenumber $k$ and frequency $\omega$ (and hence $c=\omega/k$) as fixed for a given wave motion, and determine the qualitative features of the motion which results from varying  the vorticity $\gamma_1$, and the nondimensional parameter $A$ {(see Remark \ref{A-pos-rem} and Lemma \ref{ALemma})}. (Of course,  for a given coupled-wave motion, $\gamma_1$ and $A$ are not free-parameters, and neither are $k$ and $\omega$ fixed but, rather, their values are interlinked and specified by the dispersion relation \eqref{PressInt2}.) Using this approach we obtain qualitative descriptions of a variety of  fluid motions that are prescribed by \eqref{Upper-DS0} which exhibit disparate qualitative features for various values of the parameters $A$ and $\gamma_1$.

\subsection{Lower fluid layer}
Passing to the moving frame by way of the coordinate transformation
\begin{equation}\label{CTransf1}
X(t)=kx(t)-\omega t \quad\text{and}\quad Y(t)=k(y(t)+h),  
\end{equation}
the system is transformed to a nonlinear autonomous dynamical system
\begin{equation}\label{Lower-DS}
 \begin{cases} \dot X = M\cosh Y\cos X - \omega\\
\dot Y  = M\sinh Y \sin X, 
\end{cases}
\end{equation}
where \begin{equation}\label{M}
M:=\frac{a k \omega }{\sinh (kh)}=\frac{a}{h}\cdot \frac{kh}{\sinh (kh)}\cdot \omega\ll \omega, \end{equation}
since $\omega>0$ by assumption,  and $s<\sinh(s)$ for $s>0$ and  $a/h = \mathcal O(\epsilon) \ll1$ in the linear wave regime. The autonomous system \eqref{Lower-DS} meets standard regularity assumptions for the uniqueness of the Cauchy problem \cite{Meiss}, therefore its trajectories do not intersect.
 We note that \eqref{Lower-DS} matches the dynamical system  for the lower layer in the irrotational two-layer problem that was studied in \cite{HV-JDE},  and the analysis of the phase portrait is identical. We include here the salient points for completeness.
The dynamical system \eqref{Lower-DS} has the standard (canonical) form  \cite{Meiss} of Hamilton's equations
\[
\dot X= \frac{\partial H}{\partial Y}, \qquad \dot Y = -\frac{\partial H}{\partial X}
\] 
 for the Hamiltonian function $H(X,Y)=M \sinh{Y}\cos X -\omega Y$ (where we reuse the $H$ notation for convenience: this is not the same as the Hamiltonian functional  defined in \eqref{Hexp}). Hence, if  $\left(X(t),Y(t)\right)$ is a trajectory of \eqref{Lower-DS}, then we have $\frac{d }{dt} H (X(t),Y(t))=\frac{\partial H}{\partial X} \dot X + \frac{\partial H}{\partial Y} \dot Y =-\dot Y \dot X + \dot X \dot Y=0$, and so singular points of \eqref{Lower-DS}  correspond to critical points of $H$. The nature of these points can be determined from the Hessian of $H$, and the associated Morse index, which corresponds to the number of negative eigenvalues of the Hessian. Morse's Lemma \cite{MT} implies that points with Morse index 1 are saddle points, whereas centres have Morse indices 0 and 2.

\begin{figure}[h!]
\begin{center}
 \resizebox{.7\textwidth}{!}{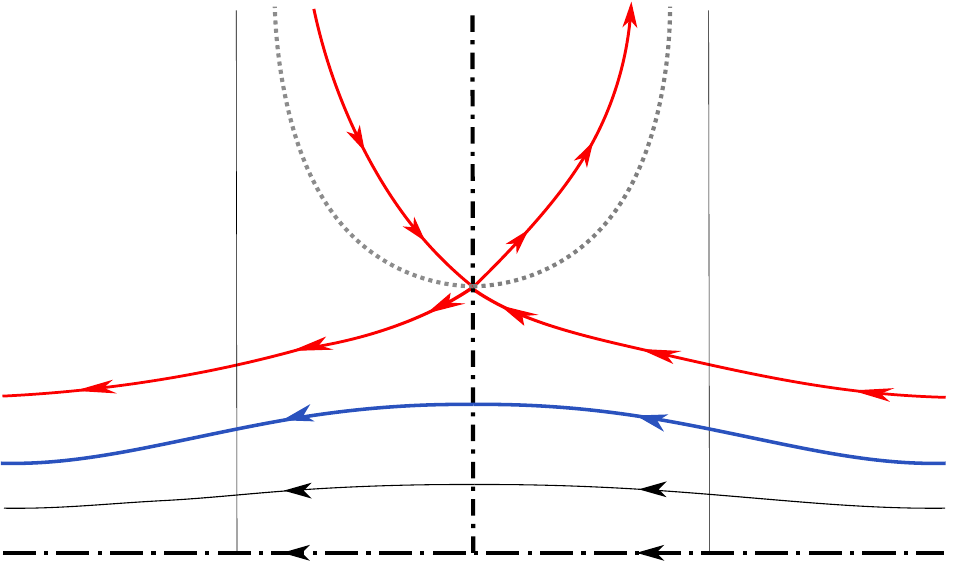}
\vspace*{5mm} 
 \caption[Cap]{
Phase portrait for the lower-fluid layer. The dotted grey line \tikz[baseline=-0.5ex]\draw[color=gray, very thick, densely dotted] (0,0) -- (.4,0); represents the $\infty-$isocline, with the dotted-dashed lines \tikz[baseline=-0.5ex]\draw[thick,dash dot] (0,0) -- (.5,0);  representing the $0-$isoclines. The internal wave profile (\tikz[baseline=-0.5ex] \draw[color=bleudefrance, very thick] (.0,0) -- (.7,0);) with mean-water level $Y=kh$ (corresponding to $y=0$) is also illustrated.
}
\label{fig:PP-Lower}
\end{center}
\end{figure}

The coordinate transformation \eqref{CTransf1}, together with the periodicity of the dynamical system with respect to $X$, mean that it suffices to consider  $\{X: -\pi \leq  X \leq \pi\}$  and $\{Y: 0\leq Y \leq k h +\e\cos X \}$, where $\e= ak$ is the  wave steepness parameter for the internal wave. The $0-$isocline is defined to be the set where $dY/dt=0$, and the $\infty-$isocline is the set where $dX/dt=0$. Therefore the $0-$isocline is given by the lines $X=0,\pm \pi$, and the line segment $Y=0$. The $\infty-$isocline is given in the region $X\in \left(-\frac{\pi}{2},\frac{\pi}{2}\right)$ by the curve $(X,\alpha(X))$, where $\alpha(X)\in \left[Y^*,\infty\right)$ for $\cosh\left(Y^*\right)={\omega}/{M}$, and $\alpha$ is defined as follows: on $\left[0,\frac{\pi}{2}\right)$ we set $\alpha$ to be the inverse of the function $Y\mapsto \arccos\left(\frac{\omega}{M\cosh(Y)}\right)$ defined on $\left[Y^*,\infty\right)$, and extend it by mirror symmetry to the interval $\left(-\frac{\pi}{2},\frac{\pi}{2}\right)$.
 Since  $\frac{\omega}{M\cosh (Y)}\leq 1$ for $Y\geq Y^*$, it follows that $\alpha$ is well-defined; furthermore the even function $\alpha$ is smooth, it takes on its infimum $Y^*$ at $X=0$, and satisfies the limiting condition $\lim_{X\rightarrow \pm \frac{\pi}{2}}\alpha(X)=\infty$.
The only singular point of the dynamical system \eqref{Lower-DS}, for positive $Y$, is $Q=(0,Y^*)$,  and the Hessian of $H$ at $Q$ is \[
\left(\begin{array}{cc}-M\sinh(Y^*) & 0 \\ 0 & M\sinh(Y^*) \end{array}  \right),
\] and so Morse's Lemma implies that $Q$ is a saddle point.
A phase portrait of the dynamical system is given in Figure \ref{fig:PP-Lower}.  
\subsection{Upper fluid layer}
When considering the fluid motion in the upper layer it is expedient to introduce the change of variables
\begin{equation}\label{CTransf2}
X(t)=kx(t)-\omega t\quad\text{and}\quad Y_1(t) = k(h_1 -y(t)),   
\end{equation}
with these new coordinates transforming the horizontal variable to the moving frame, and reversing the orientation of the vertical coordinates. The system \eqref{Upper-DS0} is thus transformed to the nonlinear autonomous  dynamical system
\begin{equation}\label{Upper-DS}
 \begin{cases} \dot X = F(X,Y_1):=M_1\cos X f(Y_1)-k\cdef -\gamma_1  Y_1, \\\
\dot Y_1  = G(X,Y_1):= M_1 \sin X g(Y_1), 
\end{cases}
\end{equation}
where 
\begin{equation}\label{M1}
M_1=a_1 k^2\cdef=\mathfrak e_1 k \cdef, \end{equation}
with $|M_1|=|\mathfrak e_1 k \cdef|\ll k|\cdef|$ (noting that $\mathfrak e_1=a_1k\ll1$)  
and
 \[
f(Y_1):=  A \cosh Y_1 - \sinh Y_1, \quad g( Y_1 ) :=  A \sinh Y_1 - \cosh Y_1.
\]
 For the remainder of the Section we assume that $a_1>0$,  without loss of generality, since shifting $X$ by $\frac{\pi}{2}$ maps \eqref{Upper-DS} to an identical system with $-a_1$ in place of $a_1$.
   The dynamical system \eqref{Upper-DS} is periodic in $X$, so we may restrict our attention to the interval  $X \in [-\pi,\pi]$. Since the coordinate transformation \eqref{CTransf2} reverses the vertical orientation, the streamline for the internal wave, $kh_1 - \e \cos X$ will lie above that of the surface wave,  $-\e_1 \cos X$, in terms of $(X,Y_1)-$coordinates, and the upper fluid layer is located in the region $\{Y_1: -\e_1\cos X \leq Y_1 \leq k h_1 - \e \cos X \}$.  As $\dot X$ and $\dot Y_1$ are given by even and odd expressions in $X$, respectively,  via \eqref{Upper-DS}, it follows that the corresponding trajectories (along with the $0-$ and $\infty-$isoclines) will be symmetric with respect to the $Y_1$ axis. 
    It follows  immediately from \eqref{Upper-DS} that the points 
    \begin{equation}\label{PPM}
    P_{\pm}^{\infty}=\left(\pm \frac{\pi}{2},-\frac{k\cdef}{\gamma_1}\right)
    \end{equation} lie in the $\infty-$isocline for any configuration of the system \eqref{Upper-DS}.  If $\gamma_1\geq0$ and $\cdef>0$ (and so there is no critical layer) then the points  $P_{\pm}^{\infty}$ must always lie below the upper fluid domain (in terms of $(X,Y_1)-$coordinates), while for $\gamma_1<0$ the points $P_{\pm}^{\infty}$ lie above the upper fluid domain (in terms of $(X,Y_1)-$coordinates) if $c/|\gamma_1|>h_1$.
The dynamical system \eqref{Upper-DS}  is also a Hamiltonian system, with Hamiltonian function $H_1(X,Y_1) = M_1\cos{X}g(Y_1)-\frac{\gamma_1} 2 Y_1^2 - k\cdef Y_1$ since
\begin{equation}\label{HamEq}
\dot X=F(X,Y_1)= \frac{\partial H_1}{\partial Y_1}, \qquad \dot Y_1 =G(X,Y_1)= -\frac{\partial H_1}{\partial X},
\end{equation}
where $H_1$ is constant along trajectories of \eqref{Upper-DS}, and singular points of \eqref{Upper-DS} correspond to critical points of $H_1(X,Y_1)$.
The Hessian matrix of  $H_1$ is
\begin{align} 
&D^2 H_1 (X,Y_1) =\begin{pmatrix}
&-M_1\cos X g(Y_1) & -M_1\sin X f(Y_1) \\
&-M_1 \sin X f(Y_1)&M_1\cos X g(Y_1) -\gamma_1 
\end{pmatrix}.\label{Hessian_H1_fg}
\end{align} 
    
\subsection{The case $\mathbf{A=1}$}\label{PP-A=1}
When $A=1$, \eqref{Upper-DS} reduces to
\begin{equation}\label{Upper-DS-A1} 
 \begin{cases} \dot X =F(X,Y_1)= M_1\cos X e^{-Y_1}-k\cdef- \gamma_1 Y_1\\\
\dot Y_1  =G(X,Y_1)= -M_1 \sin Xe^{-Y_1}.
\end{cases}
\end{equation}
The $0-$isocline for \eqref{Upper-DS-A1} (the set where $\dot Y_1=G(X,Y_1)=0$) consists of the lines $X=0,\pm \pi$. For $\cdef>0$ we have {$\dot Y_1 <0$} for $\{X:0<X<\pi\},$ while $\dot Y_1>0$ for  $\{X:0<X<\pi\}$ when $\cdef<0$ (in which case a critical layer also occurs).
The $\infty-$isocline is defined to be the set of $(X,Y_1)$ for which
\begin{equation}\label{infiso1}
\dot X=F ( X, Y_1 ) = 0 \iff  \frac { k\cdef } { \gamma_1 } + Y_1   =  \frac{ M_1 \cos X e^{-Y_1}} { \gamma_1}.
\end{equation}
Furthermore, for $A=1$ expression \eqref{A} becomes
\begin{equation}\label{Cond1}
\frac a {a_1} \frac c {\cdef} = e^{-kh_1} \Longleftrightarrow  \gamma_1 = \frac{c}{h_1}\left(1-\frac{a}{a_1}e^{k h_1}\right)\Longleftrightarrow M_1 e^{-kh_1} = \omega \e.
\end{equation}
From \eqref{Cond1} we can immediately infer a number of features of coupled surface and internal waves. For instance, ${a_1}\geq {a}e^{kh_1}\Leftrightarrow \gamma_1\geq 0$. Furthermore out-of-phase waves only occur if $\frac{c}{\cdef}<0$, which implies that the vorticity $\gamma_1>0$ is positive, and a critical layer occurs. In the shallow-water (or long-wave) regime, $kh_1\ll 1$ and $e^{-kh_1}\approx 1$. 

Before constructing some phase portraits of  \eqref{Upper-DS-A1}, we recall that the Lambert $W$ function is defined \cite{CW} to be the inverse of the function $xe^x$, that is, $W(xe^{x})=x$. It is a multivalued function: if $b>0$, then the equation $xe^x=b$ has one solution  given by $x=W(b),$ with $W(\cdot)$ standing for the principal branch of the Lambert $W$ function. When $b\in[-\frac{ 1}{e},0)$, there are two solutions, $x_1,x_2,$ with $x_1=W(b)$ and $x_2=W_{-1}(b)$, with $W_{-1}$ being the second real-valued branch of the Lambert $W$ function. The function $W:[-\frac 1 e,\infty)\to 
[-1, \infty)$ is increasing in its domain, whereas $W_{-1}:[-\frac 1 e,0)\to 
(-\infty,-1]$ is decreasing. Also, $W(-\frac 1 e)=W_{-1}(-\frac 1 e)=-1$. 
In summary, we have
\begin{equation}\label{Lamb}
x e^x =b \iff x= \begin{cases} W(b), \quad &b\geq 0\\W(b)\text{ or } W_{-1}(b),\quad&b\in[-\frac 1 e, 0).
\end{cases}
\end{equation}
\vspace{-2em}
\begin{figure}[H]
\begin{center}
\resizebox{.5\textwidth}{!}{
\begingroup%
  \makeatletter%
  \providecommand\color[2][]{%
    \errmessage{(Inkscape) Color is used for the text in Inkscape, but the package 'color.sty' is not loaded}%
    \renewcommand\color[2][]{}%
  }%
  \providecommand\transparent[1]{%
    \errmessage{(Inkscape) Transparency is used (non-zero) for the text in Inkscape, but the package 'transparent.sty' is not loaded}%
    \renewcommand\transparent[1]{}%
  }%
  \providecommand\rotatebox[2]{#2}%
  \newcommand*\fsize{\dimexpr\f@size pt\relax}%
  \newcommand*\lineheight[1]{\fontsize{\fsize}{#1\fsize}\selectfont}%
  \ifx\svgwidth\undefined%
    \setlength{\unitlength}{414.47791392bp}%
    \ifx\svgscale\undefined%
      \relax%
    \else%
      \setlength{\unitlength}{\unitlength * \real{\svgscale}}%
    \fi%
  \else%
    \setlength{\unitlength}{\svgwidth}%
  \fi%
  \global\let\svgwidth\undefined%
  \global\let\svgscale\undefined%
  \makeatother%
  \begin{picture}(1,0.50362314)%
    \lineheight{1}%
    \setlength\tabcolsep{0pt}%
    \put(0.09574281,0.07212358){\makebox(0,0)[lt]{\lineheight{1.25}\smash{\begin{tabular}[t]{l}\tb{\Large{$W_{-1}(b)$}}\end{tabular}}}}%
    \put(0,0){\includegraphics[width=\unitlength,page=1]{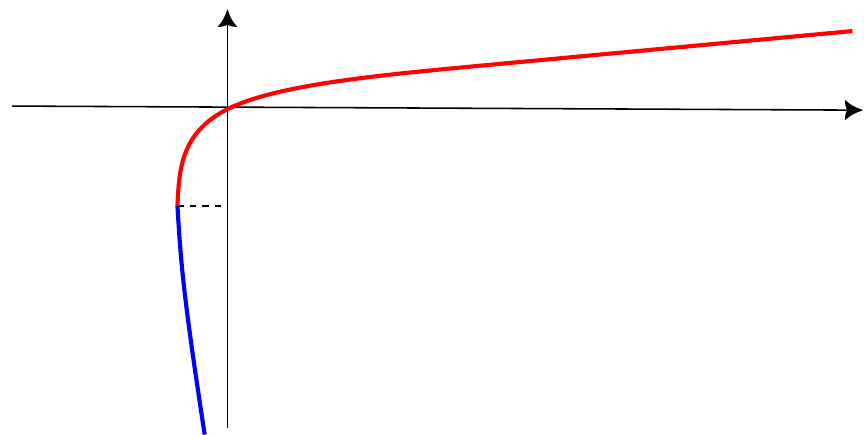}}%
    \put(1.10769169,-0.31875374){\color[rgb]{0,0,0}\makebox(0,0)[lt]{\lineheight{1.25}\smash{\begin{tabular}[t]{l}x\end{tabular}}}}%
    \put(1.00279887,0.36936782){\makebox(0,0)[lt]{\lineheight{0}\smash{\begin{tabular}[t]{l}\Large{$x$}\end{tabular}}}}%
    \put(0.2742881,0.47630708){\makebox(0,0)[lt]{\lineheight{0}\smash{\begin{tabular}[t]{l}\Large{$y$}\end{tabular}}}}%
    \put(0.2746983,0.25734976){\makebox(0,0)[lt]{\lineheight{0}\smash{\begin{tabular}[t]{l}\Large{$-1$}\end{tabular}}}}%
    \put(0.4389881,0.43833773){\makebox(0,0)[lt]{\lineheight{1.25}\smash{\begin{tabular}[t]{l}\tr{\Large{$W(b)$}}\end{tabular}}}}%
    \put(0,0){\includegraphics[width=\unitlength,page=2]{LambertW.pdf}}%
    \put(0.1734102,0.39996208){\makebox(0,0)[lt]{\lineheight{0}\smash{\begin{tabular}[t]{l}\Large{$-\frac{1}{e}$}\end{tabular}}}}%
  \end{picture}%
\endgroup%
}
\caption{The two real-valued branches of the Lambert $W$ function.}
\label{fig:lw}
\end{center}
\end{figure}
\subsubsection{$A=1$ and $\cdef>0$ (no critical layers)}
\begin{proposition}\label{P:A=1-a}
For $A=1$, let $\gamma_1> 0$ be such that $\cdef>0$.
Then the $\infty-$isocline of the dynamical system \eqref{Upper-DS} for  $X\in[0, \frac {\pi} 2]$  is given by the graph of a decreasing  function of $X$, and there exists a saddle point located at $Q_0^s=\left(0,W \left( \frac{M_1 } {\gamma_1} e^{ \frac {k \cdef } {\gamma_1}}\right) -\frac {k \cdef } {\gamma_1}\right)$,
where $W \left( \frac{M_1 } {\gamma_1} e^{ \frac {k \cdef } {\gamma_1}}\right) -\frac {k \cdef } {\gamma_1}<0$.
 In addition, $\dot X<0$ holds  for streamlines lying above the horizontal line $Y_1 = W \left( \frac{M_1 } {\gamma_1} e^{ \frac {k \cdef } {\gamma_1}}\right) -\frac {k \cdef } {\gamma_1}$.
 Furthermore, if the condition 
 \begin{equation}\label{Cond:A=1-a}
 \e_1\frac {\omega}{\gamma_1}e^{\frac{\omega}{\gamma_1}}< e^{-1}
 \end{equation} holds true, then the $\infty-$isocline of the dynamical system \eqref{Upper-DS} in $(\frac {\pi} 2,  \pi]$  consists of the graphs of a decreasing and an increasing function of $X$.  In particular,  there exist $2$  additional singular points $Q_{\pi}^{c},Q_{\pi}^{s}$ lying on the vertical line $X= \pi$ which correspond to a centre, and a saddle, respectively. Moreover,  $\dot X>0$ for $(X,Y_1)$ between these two graphs.
\end{proposition}

\begin{figure}[H]
\begin{center}
\resizebox{.7\textwidth}{!}{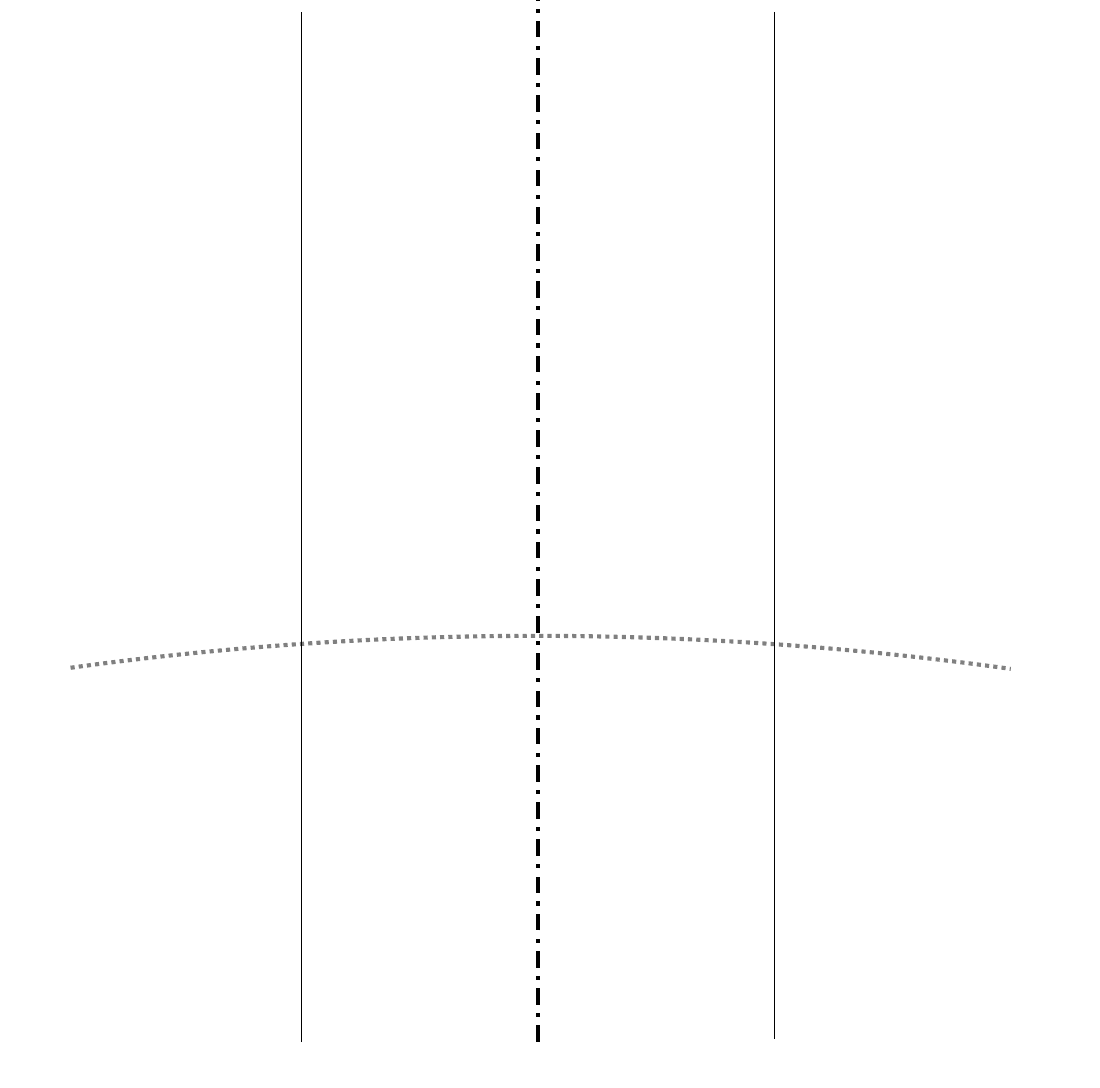}
\caption[C]{Phase portrait for $A=1$, $\gamma_1>0$, $\cdef>0$, and where \eqref{Cond:A=1-a} holds. 
The dotted grey lines \tikz[baseline=-0.5ex] \draw[color=gray, very thick, densely dotted] (.0,0) -- (.4,0) ; represent the $\infty-$isoclines, while the dotted-dashed lines \tikz[baseline=-0.5ex] \draw[thick,dash dot] (.0,0) -- (.5,0) ;  represent the $0-$isoclines.  The {surface wave} profile (\tikz[baseline=-0.5ex] \draw[color=azure, very thick] (.0,0) -- (.7,0);) has mean-water level $Y_1=0$, corresponding to $y=h_1$. The internal wave profile (\tikz[baseline=-0.5ex] \draw[color=bleudefrance, very thick] (.0,0) -- (.7,0);) has mean water level $Y=kh_1$, corresponding to $y=0$.
 Note that the streamline representing the surface wave lies {\em beneath} that of the internal wave in terms of the $(X,Y_1)-$coordinates used for this phase portrait.
}
\label{fig:PP-A=1-a}
\end{center}
\end{figure}
\begin{proof}
First observe that for $\gamma_1>0$ and $\cdef>0$, \eqref{Cond1} implies that $0<a<a_1$, and so the surface wave must be larger than the internal wave in this regime.
For $X\in[0, \frac {\pi} 2]$, where $\cos X \geq 0$, we observe  that the condition \eqref{infiso1}  can be expressed in terms of the Lambert $W$ function \eqref{Lamb} as
\begin{equation*}
F ( X, Y_1 ) = 0 \iff  \frac { k\cdef } { \gamma_1 } + Y_1   =  W \left( e^{\frac { k\cdef } { \gamma_1 }  }  \frac{ M_1 \cos X} { \gamma_1 }\right).
\end{equation*}  Hence the dynamical system \eqref{Upper-DS-A1} has a singular point at $Q_0^s=(0,\ym(0))$, where $\ym(0)= W \left( \frac{M_1 } {\gamma_1} e^{ \frac {k \cdef} {\gamma_1}}\right) -\frac {k \cdef } {\gamma_1}<0$, since $M_1=\mathfrak e_1 k\cdef\ll k\cdef$ and $W$ is an increasing function on the positive real-line.
This non-hyperbolic singular point can be  characterised by exploiting the Hamiltonian structure of \eqref{Upper-DS} and Morse theory. The Hessian at $Q_0^s$ is 
\[
D^2 H_1 (0,\ym(0) ) =\begin{pmatrix}
&  M_1  e^{-\ym(0) } &0\\
&0 & {-M_1  e^{- \ym(0) }-\gamma_1}
\end{pmatrix},
\] 
and the diagonal terms have opposite signs, hence $Q_0^s$ is a saddle point lying on the intersection of four separatrices, which are the subsets of the phase space $(X,Y_1)$ for which $H_1(X,Y_1)=H_1(0, \ym(0))$. We note that $F(X,0)=M_1\cos X - k\cdef=\e_1 k\cdef 
\cos X - k\cdef<0$, due to \eqref{M1}, and ${F} (X,kh_1)=M_1 \cos X e^{-kh_1}-\omega=\e \omega \cos X -\omega<0$, due to \eqref{M} and  \eqref{Cond1}. 

To establish the  second claim, if the assumed condition \eqref{Cond:A=1-a} on the wave-steepness holds then $-\frac 1 e <\frac {M_1 \cos {X}} {\gamma_1}e^{\frac{k\cdef}{\gamma_1}}< 0$ when $X\in(\frac{\pi}{2},\pi]$.  The $\infty-$isocline then consists of two different sets of points $(X,\ym(X))$ and $(X,\yl(X))$ which are formed by the respective branches of the Lambert $W$ function \eqref{Lamb}, namely:
\[    \ym(X)  =  W \left( \frac{M_1 \cos X} {\gamma_1} e^{ \frac {k \cdef } {\gamma_1}}\right) -\frac {k \cdef } {\gamma_1},
\qquad 
    \yl(X) =  W_{-1} \left( \frac{M_1 \cos X} {\gamma_1} e^{ \frac {k \cdef} {\gamma_1}}\right) -\frac {k \cdef } {\gamma_1}.
\]
The monotonicity properties of the Lambert $W$ function, and the fact that $X\in(\frac{\pi}{2},\pi]$, together yield that the former branch corresponds to the graph of a decreasing function, and the latter to that of an increasing one. Recalling the form of the $0-$isocline, there are singular points $Q_{\pi}^{c}=( \pi,\ym(\pi))$ and  $Q_{\pi}^{s}=( \pi,\yl(\pi))$ where
\[
 \ym(\pi) = W \left( -\frac{M_1 } {\gamma_1} e^{ \frac {k \cdef } {\gamma_1}}\right) -\frac {k \cdef } {\gamma_1}
\quad
\text{and}
\quad \yl(\pi) = W_{-1} \left( -\frac{M_1 } {\gamma_1} e^{ \frac {k \cdef } {\gamma_1}}\right) -\frac {k \cdef} {\gamma_1}.
 \]
For the point $Q_{\pi}^{c}$ the Hessian is
\[
D^2 H_1 (\pi,\ym(\pi)) =\begin{pmatrix}
&-M_1  e^{-\ym(\pi)} &0\\
&0 &M_1  e^{-\ym(\pi)} {-\gamma_1}
\end{pmatrix},\] 
while the Hessian at $Q_{\pi}^{s}$ is given by
\[
D^2 H_1 (\pi,\yl(\pi)) =\begin{pmatrix}
&-M_1  e^{-\yl(\pi)} &0\\
&0 &M_1  e^{-\yl(\pi)}-\gamma_1
\end{pmatrix}.\] 
 For fixed $X\in (-\frac{\pi}{2},\pi]$, {$F(X,\cdot)$} is increasing for  $Y_1<\ln\left(-\frac{M_1 \cos X}{\gamma_1}\right),$ and decreasing for $Y_1$ above this value. Note that $\ln\left(-\frac{M_1 \cos X}{\gamma_1}\right)<-1-\frac{k\cdef}{\gamma_1}<-\e_1$, since we are assuming that $\frac{M_1}{\gamma_1}e^{\frac{k\cdef}{\gamma_1}}<\frac{1}{e}$. These monotonicity properties imply that
\[
M_1  e^{-\ym(\pi)}-\gamma_1=\frac {\partial F}{\partial Y_1}(\pi,\ym(\pi))<0, \quad 
M_1  e^{-\yl(\pi)}-\gamma_1=\frac {\partial F}{\partial Y_1}(\pi,\yl(\pi))>0.
\]
Since $M_1>0$,   Morse theory implies that $Q_{\pi}^{c}=(\pi,\ym(\pi))$ corresponds to a centre, as it has Morse index 2, and $Q_{\pi}^{s}=(\pi,\yl(\pi))$ to a saddle, as it has Morse index 1. Finally, the fact that {$F(X,\cdot)$} changes monotonicity once in $Y_1$, for fixed $X$, first being increasing, shows that $\dot X>0$ between the two branches of the $\infty-$isocline.
 \end{proof}
\begin{remark}
In order for all  singular points for the dynamical system \eqref{Upper-DS} considered in Proposition \ref{P:A=1-a} to lie below the fluid domain we need
\begin{equation*}
W \left( \frac{M_1 } {\gamma_1} e^{ \frac {k \cdef} {\gamma_1}}\right) -\frac {k \cdef } {\gamma_1}=
W \left( \frac{\e \omega} {\gamma_1} e^{ \frac {\omega} {\gamma_1}}\right) -\frac{\omega}{\gamma_1} + kh_1<-\e_1.
\end{equation*}
If this holds true, then $\dot X<0$ along all streamlines.
\end{remark}
\begin{remark}
Two saddle-centre bifurcations occur when $\frac{M_1}{\gamma_1} e^{\frac{k\cdef}{\gamma_1}}$ crosses the value $-\frac{1}{e}$: there is a transition from $1$ to $5$ critical points for the dynamical system.
\end{remark}

\begin{proposition}\label{P:A=1-b}
For $A=1$, let $\gamma_1<0$ (and so $\cdef>0$).
Then the $\infty-$isocline of the dynamical system \eqref{Upper-DS} in  $[\frac{\pi}{2}, \pi]$  is given by the graph of an increasing  function of $X$, and there exists a saddle point located at $Q_{\pi}^s=\left(\pi,W \left( -\frac{M_1 } {\gamma_1} e^{ \frac {k \cdef } {\gamma_1}}\right) -\frac {k \cdef } {\gamma_1}\right)$, where $W \left( -\frac{M_1 } {\gamma_1} e^{ \frac {k \cdef } {\gamma_1}}\right) -\frac {k \cdef } {\gamma_1}>0$. In addition, $\dot X>0$ for $(X,Y_1)$ above this graph.
Furthermore, if the condition 
\begin{equation}
\label{Cond:A=1-b}
\e_1 e^{\e_1}<1
\end{equation}  holds true, then the $\infty-$isocline for the dynamical system \eqref{Upper-DS} in $[0,\frac {\pi} 2)$  consists of the graphs of a decreasing and an increasing function of $X$.  In particular,  there exist $2$ additional singular points $Q_{0}^{c},Q_{0}^{s}$ lying on the vertical line $X=0$ which correspond to a centre, and a saddle, respectively. Moreover,  $\dot X<0$ for $(X,Y_1)$ between the two graphs.
\end{proposition}
  \begin{figure}[H]
\begin{center}
\resizebox{.7\textwidth}{!}{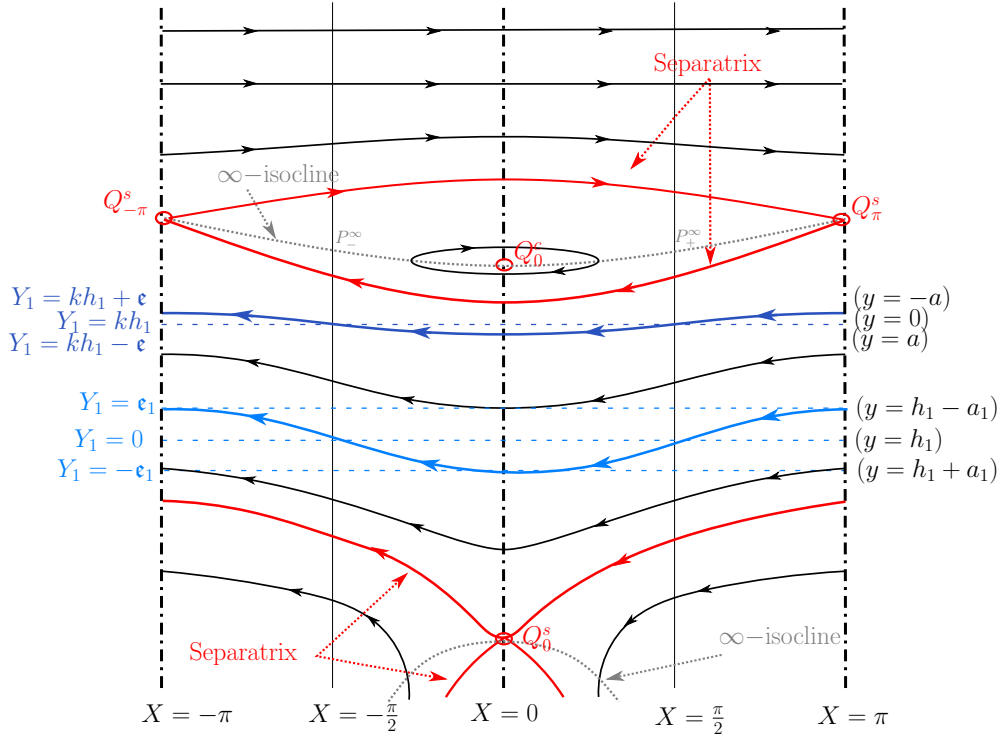}
\caption{Phase portrait for $A=1$, $\cdef>0$, $\gamma_1<0$, where \eqref{Cond:A=1-b} holds. A similar legend applies as in Figure \ref{fig:PP-A=1-a}.}
\label{fig:PP-A=1-b}
\end{center}
\end{figure}
\begin{proof}
Note that \eqref{Cond1} implies that $a>a_1e^{-kh_1}>0$ for $\gamma_1<0$, and hence the internal and surface waves must be in-phase in this regime. Since many elements of the proof follow that of Proposition \ref{P:A=1-a} (upon swapping the intervals $[-\frac {\pi} 2, \frac {\pi} 2]$ and $[-\pi, \pi]\setminus[-\frac {\pi} 2, \frac {\pi} 2]$) we just provide the essential details. 
When $X\in [\frac{\pi}{2}, \pi]$, condition \eqref{infiso1} prescribing the $\infty-$isocline can be reexpressed as
\[
 F(X,Y_1)= 0 \iff Y_1 = W \left(\frac{M_1 \cos X}{\gamma_1}e^{\frac{k\cdef}{\gamma_1}} \right)-\frac{k\cdef}{\gamma_1} ,
\] since $\gamma_1<0$.  We note that $\frac{\partial F}{\partial Y_1}>0$ for points along the $\infty-$isocline when  $X\in [\frac{\pi}{2}, \pi]$. The $\infty-$isocline is given by the graph of the increasing function  $W\left(\frac{M_1 \cos X}{\gamma_1}e^{\frac{k\cdef}{\gamma_1}}\right)-\frac{k\cdef}{\gamma_1}$ in this case. Recalling the form of the $0-$isocline, there is a singular point at $Q_{\pi}^s=(\pi,\ym(\pi) )$, where $\ym(\pi)  = W \left(- \frac{M_1 } {\gamma_1} e^{ \frac {k \cdef } {\gamma_1}}\right) -\frac {k \cdef } {\gamma_1}>0$. The Hessian  at $Q_{\pi}^s$ is given by
\[
D^2 H_1 (\pi,\ym(\pi) ) =\begin{pmatrix}
&M_1  e^{-\ym(\pi) } &0\\
&0 &-M_1  e^{-\ym(\pi) }-\gamma_1
\end{pmatrix}.\]
Since $M_1$ is positive, and $\frac{\partial F}{\partial Y_1}(\pi, \ym(\pi))=-M_1  e^{-\ym(\pi) }-\gamma_1>0$ due to the monotonicity properties of $F(X,\cdot)$ along the $\infty-$isocline for  $X\in [\frac{\pi}{2}, \pi]$, this point has Morse index 1 and hence $Q_{\pi}^s$ is a saddle.

If \eqref{Cond:A=1-b} holds, note that $\e_1 e^{\e_1}<1$ implies that $\frac{M_1}{\gamma_1}e^{\frac{k\cdef}{\gamma_1}}>-\frac 1 e$ since
  \begin{align}\label{Prop-A_1_2_eq}
 \e_1 e^{\e_1}<1\implies \frac{M_1}{\gamma_1} e^{\e_1}>\frac{k\cdef}{\gamma_1}
 \\
 \nonumber \implies \frac{M_1}{\gamma_1}e^{\e_1}>\frac{k\cdef}{\gamma_1}-\e_1\implies \frac{M_1}{\gamma_1}e^{\frac{k\cdef}{\gamma_1}}
>\left(\frac{k\cdef}{\gamma_1}-\e_1\right)e^{\frac{k\cdef}{\gamma_1}-\e_1}>-\frac{1}{e},
 \end{align}
 where $M_1=\e_1 k\cdef$ and the last step follows from observing that the function $se^s$ has a minimum value at $s=-1$.  Thus, since $X\in[0, \frac {\pi} 2)$,  relation \eqref{Lamb} implies that the equation $F(X,Y_1)=0$ has two solutions in $Y_1$, for fixed $X$ in this range, and so the $\infty-$isocline consists of the graphs of the functions 
 \[
 \ym (X)=W\left(\frac{M_1 \cos X}{\gamma_1}e^{\frac{k\cdef}{\gamma_1}}\right)-\frac{k\cdef}{\gamma_1} \ \mbox{ and }\ \yl(X)=W_{-1}(\frac{M_1 \cos X}{\gamma_1}e^{\frac{k\cdef}{\gamma_1}})-\frac{k\cdef}{\gamma_1}.
 \] 
 Recalling that the $0-$isocline consists of the lines $\{X=0,\pm \pi\}$,  there exist additional singular points $Q_0^s=(0,\yl(0)), Q_0^c=(0,\ym(0))$, where
\[
 \ym(0) = W \left( \frac{M_1 } {\gamma_1} e^{ \frac {k \cdef } {\gamma_1}}\right) -\frac {k \cdef } {\gamma_1}>0,
\quad
\quad
\quad \yl(0) = W_{-1} \left( \frac{M_1 } {\gamma_1} e^{ \frac {k \cdef } {\gamma_1}}\right) -\frac {k \cdef } {\gamma_1}<0.
\]
Since $M_1>0$, evaluating the Hessian $D^2 H_1$ \eqref{Hessian_H1_fg} implies by Morse theory that  $Q_0^c=(0,\ym(0))$ corresponds to a centre, since $\frac {\partial F }{\partial Y_1}(0,\ym(0))>0$, whereas $(0,\yl(0))$  is a saddle point since $\frac {\partial  F}{\partial Y_1}(0,\yl(0))<0$.  
\end{proof}
\begin{remark}
In order for all  singular points for the dynamical system \eqref{Upper-DS} considered in Proposition \ref{P:A=1-b} to lie outside the fluid domain we require 
\begin{equation*}
\e<W \left( \frac{\e \omega} {\gamma_1} e^{ \frac {\omega} {\gamma_1}}\right) -\frac{\omega}{\gamma_1}
\end{equation*}
in order to to ensure that $Q_{\pi}^s$ lies above $Y_1=kh_1+\e$. Note that it follows from the monotonicity of $W_{-1}$, and \eqref{Prop-A_1_2_eq}, that $Q_{0}^s$ will lie below $Y_1=\e_1$ since
  $W_{-1}\left(\frac{M_1}{\gamma_1}e^{\frac{k\cdef}{\gamma_1}}\right)<\frac{k\cdef}{\gamma_1}-\e_1.$
Also $\dot X<0$ along all streamlines in the upper fluid layer.
\end{remark}

\subsubsection{$A=1$ and $\cdef<0$ (critical layer present)}
\begin{proposition}\label{P:A=1-c}
For $A=1$, let $\gamma_1>0$ be such that $\cdef<0$.
The $\infty-$isocline for the dynamical system \eqref{Upper-DS} in  $[\frac{\pi}{2},\pi]$  is given by the graph of an increasing  function of $X$, and there exists a saddle point located at $Q_{\pi}^s=\left(\pi,W \left( -\frac{M_1 } {\gamma_1} e^{ \frac {k \cdef } {\gamma_1}}\right) -\frac {k \cdef } {\gamma_1}\right)$. In addition, $\dot X<0$    along streamlines lying  above the horizontal line $ Y_1 = W \left( -\frac{M_1 } {\gamma_1} e^{ \frac {k \cdef } {\gamma_1}}\right) -\frac {k \cdef } {\gamma_1}$.
Furthermore, if the condition 
\begin{equation}
\label{Cond:A=1-c}
\e_1 e^{\e_1}<1
\end{equation}  holds true, then the $\infty-$isocline for the dynamical system \eqref{Upper-DS} in $[0,\frac {\pi} 2)$  consists of the graphs of a decreasing and an increasing function of $X$.  In particular,  there exist $2$ additional singular points, $Q_0^s,Q_0^c$, corresponding to a saddle point and a centre respectively, lying at the vertical line $X= 0$. Moreover,  $\dot X>0$ for $(X,Y_1)$ between the two graphs.
\end{proposition}
  \begin{figure}[H]
\begin{center}
\resizebox{.7\textwidth}{!}{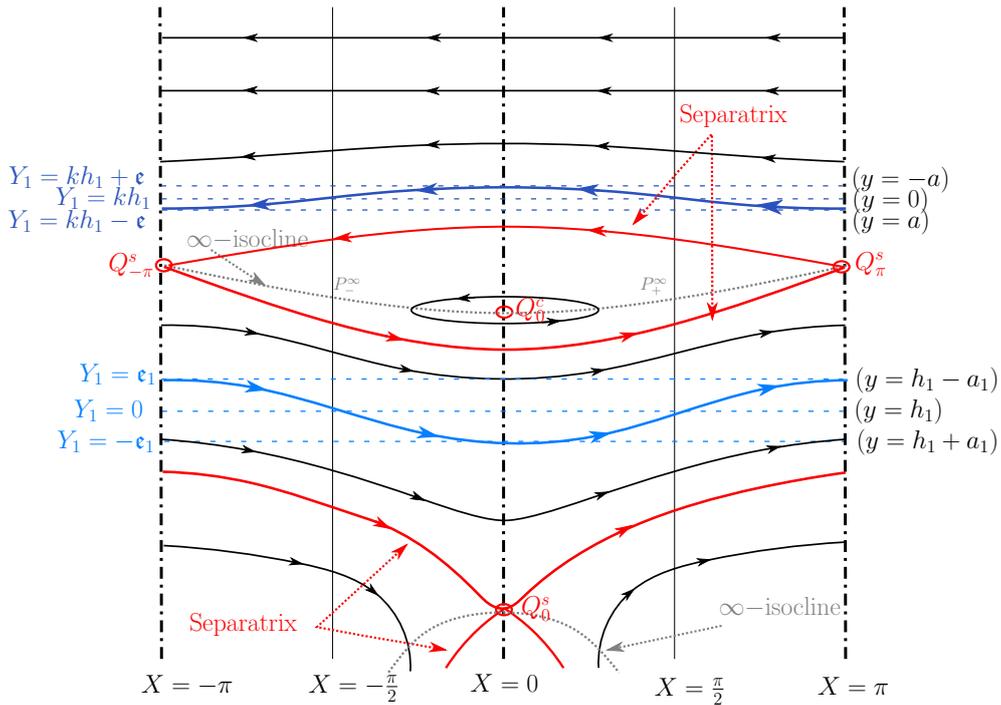}
\caption{Phase portrait for $A=1$, $\cdef<0$ ($\gamma_1>0$), where \eqref{Cond:A=1-c} holds. A similar legend applies as in Figure \ref{fig:PP-A=1-a}.}
\label{fig:PP-A=1-c}
\end{center}
\end{figure}
\begin{proof}
For $\cdef<0$, it follows from \eqref{Cond1} that $a/a_1<0$, and hence the internal and surface waves must be out of phase.
Noting that the relation \eqref{infiso1}, which defines the $\infty-$isocline, features the parameters $\cdef$ and $\gamma_1$ solely by way of their ratio. Therefore swapping the signs of each of these terms does not materially affect the considerations of Proposition \ref{P:A=1-b} relating to the structure of the $\infty-$isocline, and the existence and location of singular points. 
When $\cos X < 0$, $\frac{\partial F}{\partial Y_1}(X,\cdot)<0$;   when   $\cos X>0$, $F(X,\cdot)$ is increasing for $Y_1<\ln\left(-\frac{M_1 \cos X}{\gamma_1}\right)$, while $F(X,\cdot)$ is decreasing for $Y_1$ above this value.
Using similar reasoning to Propositions \ref{P:A=1-a} and \ref{P:A=1-b},  determining the sign of the eigenvalues for the Hessian \eqref{Hessian_H1_fg} we deduce that   $Q_0^s=(0,\yl(0))$ corresponds to a saddle point, since $\frac {\partial {F}}{\partial Y_1}(0,\yl(0))>0$ and $M_1<0$; 
$Q_0^c=(0,\ym(0))$  must be a centre, since $\frac {\partial {F}}{\partial Y_1}(0,\ym(0))<0$; while $Q_{\pm\pi}^s=(\pm \pi,\ym(\pi)) $ correspond to saddle points, since $\frac {\partial {F}}{\partial Y_1}(\pm \pi,\ym(\pi))<0$, also recalling the sign of $M_1$. 

For $\cdef<0$, the points $P_{\pm}^{\infty}=\left(\pm \frac{\pi} 2, -\frac{k\cdef}{\gamma_1}\right)$ from the $\infty-$isocline  must be located within the fluid domain, by inspection. When condition \eqref{Cond:A=1-c} holds, the saddle point $Q_0^s=(0,\yl(0))$ lies outside the fluid domain (beneath the surface wave in terms of $(X,Y_1)-$coordinates). Then the branches of the separatrices which join the points $Q_{\pm\pi}^s$ represent critical layers for fluid motion in the upper layer, demarcating a vortex centred at $Q_0^c$ which has streamlines redolent of the Kelvin's cat's eyes patterns observed in the single fluid layer model \cite{CO}. We infer from inspecting $\frac {\partial F}{\partial Y_1}$ that $\dot X>0$ for streamlines in the fluid domain which lie under the upper $\infty-$isocline, while  $\dot X<0$ for streamlines in the fluid domain which lie above the upper $\infty-$isocline.

 \end{proof}

\begin{remark}
 Suppose that \eqref{Cond:A=1-c} holds, then in order for the critical layer to be located in the fluid domain the inequalities 
	    \begin{equation}\label{Prop-A_1_3_eq}
\frac {\omega } {\gamma_1}<W \left( \frac{\e \omega } {\gamma_1} e^{ \frac {\omega } {\gamma_1}}\right)+kh_1 \quad\text{and}\quad  W \left( -\frac{\e \omega } {\gamma_1} e^{ \frac {\omega } {\gamma_1}}\right) <\frac { \omega } {\gamma_1}-\e,
\end{equation}
must hold, which offers a further condition on $\e$.
\end{remark}

\subsection{The case $A<1$}\label{PP-A<1}
As observed in Lemma \ref{ALemma}, $A$ is permitted to be negative for rotational flows (with a lower bound of $A\geq -{\gamma_1^2}/{4gk}$), which adds an additional level of complexity compared to the setting of purely irrotational fluid layers considered in \cite{HV-JDE, HV-AN}. 
From \eqref{A-alt} we observe that 
\begin{equation}\label{Cond2}
e^{-kh_1}<\frac a {a_1} \frac c {\cdef} < \mathfrak U, 
\end{equation}
where $\mathfrak U=\cosh kh_1+{\gamma_1^2}\sinh kh_1/{4gk}$.
As stated previously, we assume without loss of generality that $a_1,c>0$ in subsequent considerations.  Hence, \eqref{Cond2} implies that waves are in phase ($a/a_1>0$) if and only if $\cdef>0$ ($M_1>0$). Hence, $\cdef<0$ ($M_1<0$) implies both the existence of a critical layer, and out-of-phase waves, as we might expect from the considerations of Section \ref{PP-A=1}. In addition, when $\cdef>0$, 
 \eqref{Cond2} implies that:
\[
  \frac{c}{h_1}\left(1-\frac{a}{a_1}e^{k h_1}\right)<\gamma_1<\frac{c}{h_1}\left(1-\frac{a}{a_1} \frac {1} {\mathfrak U} \right),
\]
with $\cdef<0$ yielding the reverse inequality.  In particular, if  $\cdef>0$ then  $a<a_1\mathfrak U$ for $\gamma_1>0$, while  $a_1<{a}e^{k h_1}$ for $\gamma_1<0$, whereas  for $\cdef<0$ (and so $\gamma_1>0$) then  $a_1>{a}e^{k h_1}$.

{For the remainder of this section we assume that $|A|<1$ (the condition $\gamma_1^2\leq 4gk$ is sufficient for this to hold)}. Regarding the dynamical system \eqref{Upper-DS}, in this setting we have $g(Y_1)=A\sinh Y_1 - \cosh Y_1<0$ for all $Y_1 \in \R$. Therefore,  when $\cdef>0$, $\dot Y_1>0$ for $X\in (-\pi,0)$ and $\dot Y_1<0$ for $X\in (0,\pi)$, with the inequality signs reversed for $\cdef<0$. The $0-$isocline must consist solely of the lines $\{X=0,\pm \pi\}$.
The $\infty-$isocline for \eqref{Upper-DS} consists of the set of points for which the function $F(X,Y_1)=0$, 
and in order to determine the form of the $\infty-$isocline a detailed insight into the  monotonicity properties of { $F(X,\cdot)$ } is required, with
\begin{equation}\label{Fy}
 \frac {\partial {F}}{\partial Y_1}=M_1 \cos X g(Y_1)-\gamma_1.
 \end{equation} For $|A|<1$ we also have $\frac {\partial^3 {F}}{\partial Y_1^3}(X,Y_1)<0$ when $X\in(-\frac{\pi}{2},\frac{\pi}{2})$, and $\frac {\partial^3 {F} }{\partial Y_1^3}(X,Y_1)>0$ when $X\in[-\pi,\pi]\setminus[-\frac{\pi}{2},\frac{\pi}{2}]$, for all $Y_1$. Also $\frac {\partial^2 {F}}{\partial Y_1^2}(\pm\frac{\pi}{2},Y_1)=0$ for all $Y_1$.
For $|A|<1$ the quantity 
$
 \Adefa:=\frac{1+A}{1-A}
 $
is well-defined, and $f(\Adefa)=0$. The relation $\frac {\partial^2 {F}}{\partial Y_1^2}\left(X,\frac 1 2 \ln\left(\frac{1+A}{1-A}\right)\right)=0$ then holds true for all $X\in [-\pi,\pi]$.
 \subsubsection{$|A|<1$ and $\cdef>0$ (no critical layers)}
For $\cdef>0$ (and so $M_1>0$)  the vorticity must satisfy either $0<\gamma_1<\frac{c}{h_1},$ or $\gamma_1<0$.  If $0<\gamma_1<\frac{c}{h_1}$ and  $X\in[-\frac{\pi}{2},\frac{\pi}{2}]$, then \eqref{Fy} implies that $\frac {\partial {F}}{\partial Y_1}<0$, and hence the $\infty-$isocline is not empty by the implicit function theorem.
Analogously, for  $\gamma_1<0$ and $X\in[-\pi,\pi]\setminus(-\frac{\pi}{2},\frac{\pi}{2})$, \eqref{Fy} implies that $\frac {\partial {F}}{\partial Y_1}>0$, and again the $\infty-$isocline is not empty by the implicit function theorem.

\begin{figure}[H]
\begin{center}
\resizebox{.5\textwidth}{!}{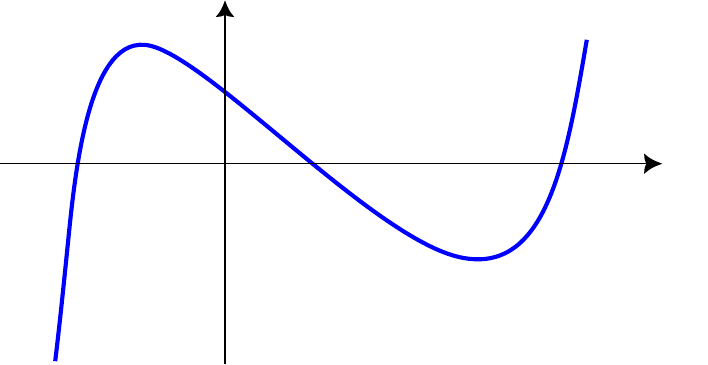}
\caption{Schematic of the function $\Fdefa (Y_1)$, when $A<1$.}
\label{fig:A_less_1}
\end{center}
\end{figure}

In the case of either $0<\gamma_1<\frac{c}{h_1}$ and a fixed  $X\in[-\pi,\pi]\setminus[-\frac{\pi}{2},\frac{\pi}{2}]$, or $\gamma_1<0$ and  $X\in(-\frac{\pi}{2},\frac{\pi}{2})$,  there is the possibility of $\frac {\partial {F}}{\partial Y_1}$ having two zeros in $Y_1$. For such fixed values of $X$, we will establish that the monotonicity behaviour of $F(X,\cdot)$ matches either that of the function $\Fdefa (Y_1)$  in Figure \ref{fig:A_less_1}, or its negative.
 Thus, to determine the structure of the $\infty-$isocline, and the existence of singular points for the dynamical system \eqref{Upper-DS}, it suffices to estimate the location of  maximum and minimum values of $F(X,\cdot)$  for $X$ in the given ranges. Relation \eqref{Fy} implies that 
 \begin{equation*}
\frac{\partial {F}}{\partial Y_1}=0\iff\sqrt{1-A^2}\,M_1\cos X \cosh Y_1^A=-\gamma_1,
\end{equation*}
where $Y_1^A:=Y_1-\frac 12 \ln \Adefa$. 
Bearing in mind $|A|<1$, we define 
\[
\mrl : = \frac{\sqrt{1-A^2}\, M_1}{\gamma_1} \quad \mbox{ and } \quad  z_1(X):=\frac{ 1}{| \mrl \cos X|}  \ \mbox{ for } X\neq \frac{\pi}{2}.
\]
Then, for $X\neq \frac{\pi}{2}$ and $|\mrl|\leq 1$,  the local extrema of ${F}(X,\cdot)$ are
\begin{align}
Y_1^{\pm}(X)=\frac 1 2 \ln\Adefa&\pm \arcosh z_1(X) \nonumber
\\
&=\ln\Adefa^{1/2}\pm \ln\left(z_1 (X)\!+\!\sqrt{z_1^2 (X)-1}\right)\! \label{Y1pm}
\end{align}
with
  \begin{align*}
  F(X,Y_1^{\pm}(X))= \mp \sqrt{ 1 - A^2 } M_1 \cos X \sinh (\arcosh z_1(X) )-k\cdef- \gamma_1 Y_1^{\pm}(X) \\
 =\gamma_1\left[\pm \left(\tanh \arcosh z_1(X) - \arcosh z_1(X)\right)-\frac{k\cdef}{\gamma_1}-\frac1 2 \ln\Adefa\right],\\
 {F}(X,Y_1^+(X))  {F}(X,Y_1^-(X)) \qquad \qquad \qquad \qquad \qquad \qquad \qquad \qquad  \quad \qquad \\
=\gamma_1^2\left[\left(\frac{k\cdef}{\gamma_1}+\frac1 2 \ln\Adefa\right)^2-\left(\tanh \arcosh z_1(X) {-} \arcosh z_1(X)\right)^2\right],
 \end{align*}
 noting that $\left[\tanh \arcosh x  - \arcosh x\right]'=-{\sqrt{x^2-1}}/{x^2},$ {for $x\geq 1.$}
Moreover,  \begin{align*}
  {F}(X,& Y_1^+(X)) {F} (X,Y_1^-(X))<0\\
  &\iff \arcosh z_1(X) {-} \tanh \arcosh z_1(X) > \left|\frac{k\cdef}{\gamma_1}+\frac 12 \ln \Adefa\right|.
   \end{align*}
   The last inequality allows us to determine  the number of possible solutions of the equation ${F}(X,\cdot)=0$, for fixed $X\neq \pm \frac{\pi}{2}$, according to the sign of the values of ${F}(X,\cdot)$ at its critical points.
In particular, if
\begin{equation}\label{Aless1cond}
\arcosh \frac{1}{|\mrl|}-\tanh \arcosh \frac{1}{|\mrl|} > \left| \frac{k\cdef}{\gamma_1} +\frac 12 \lAA\right|,
\end{equation}
then 
\begin{align*}
  F (X,& Y_1^+(X)) F(X,Y_1^-(X))<0,
   \end{align*}
   for all $X$ such that this expression is well defined.
   \begin{proposition}\label{P:Aless1-a}
For $|A|<1$,    let $\gamma_1>0$ be such that $\cdef>0$.  If $|\mrl|<1$ and \eqref{Aless1cond} holds, then the dynamical system \eqref{Upper-DS}   
 has $7$ singular points.
   \end{proposition}
       \begin{figure}[H]
\begin{center}
\resizebox{.67\textwidth}{!}{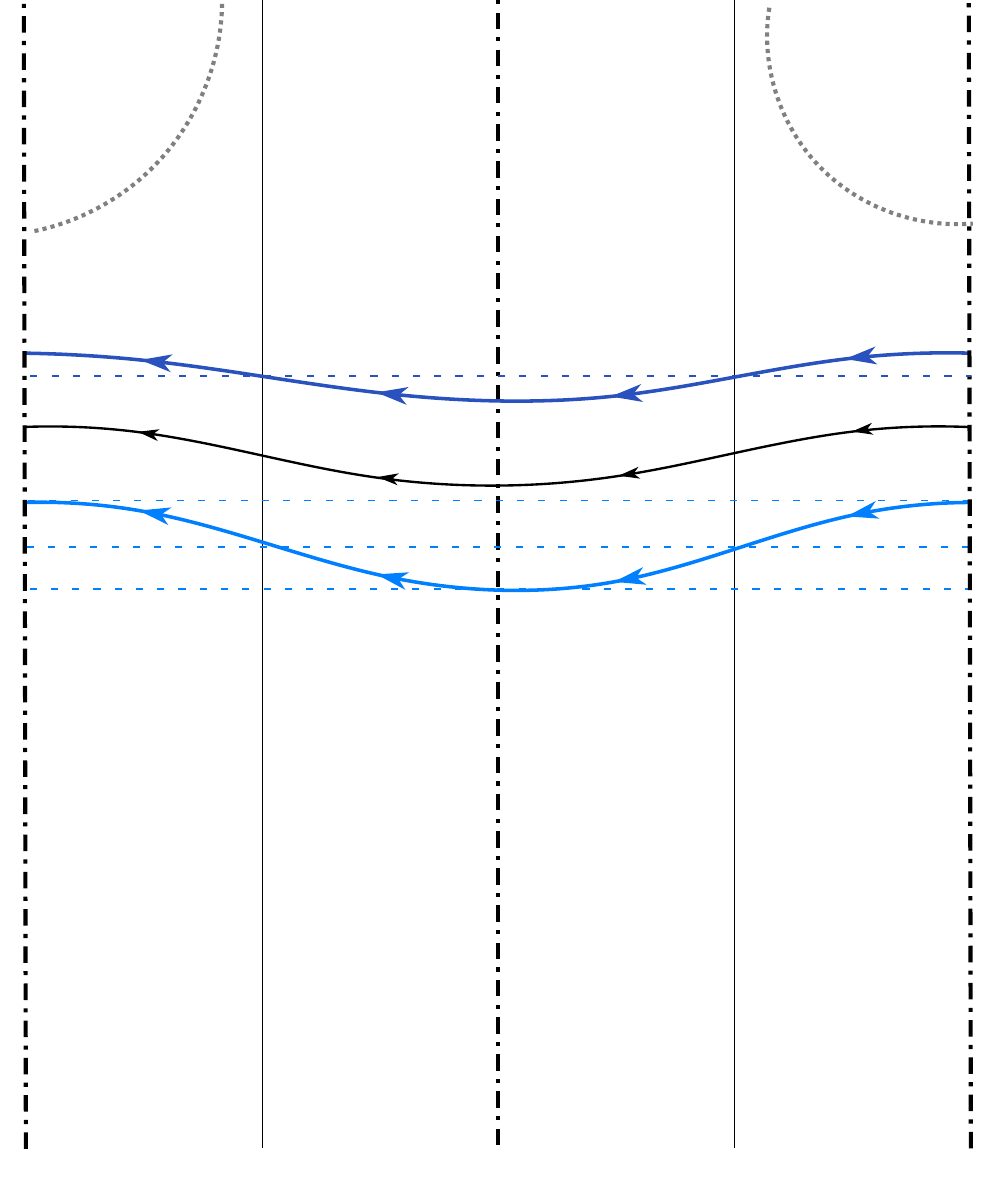}
\caption[C1]{Phase portrait for $A<1$, $\cdef>0$, $\gamma_1>0$, where \eqref{Aless1cond} holds. The dotted grey line \tikz[baseline=-0.5ex] \draw[color=gray, very thick, densely dotted] (.0,0) -- (.4,0) ; represents the $\infty-$isocline, while the dotted-dashed lines \tikz[baseline=-0.5ex] \draw[thick,dash dot] (.0,0) -- (.5,0) ;  represent the $0-$isoclines.  The {surface wave} profile (\tikz[baseline=-0.5ex] \draw[color=azure, very thick] (.0,0) -- (.7,0);) has mean-water level $Y_1=0$, corresponding to $y=h_1$. The internal wave profile (\tikz[baseline=-0.5ex] \draw[color=bleudefrance, very thick] (.0,0) -- (.7,0);) has mean water level $Y=kh_1$, corresponding to $y=0$.}
\label{fig:PP-Aless1-a}
\end{center}
\end{figure}
     \begin{proof}
        From \eqref{Fy} 
    \begin{align}  \nonumber
\frac{\partial {F}}{\partial Y_1} = M_1\cos X  g(Y_1)-\gamma_1 \qquad\qquad \qquad   \qquad\qquad \qquad  \qquad\qquad \qquad 
\\
\label{g2der}
\qquad \qquad  \qquad\qquad  =-\sqrt{1-A^2}M_1\cos X \cosh \left( Y_1 - \frac 12 \lAA \right)-\gamma_1.
\end{align}
Thus, $F(X,\cdot)$ is decreasing for $|X| \leq \frac{\pi}{2}$, with the $\infty-$isocline being non-empty, due to the implicit function theorem, and there exists a singular point along the line $X=0$, which we label $Q_0^s=(0,\ym(0))$. The analysis in the preceding section allows us to deduce that the equation $F(X,Y_1)=0$ has three roots in $Y_1$, when $\frac{\pi}{2}<|X|$, due to \eqref{Aless1cond}: $F(X,\cdot)$ shares similar monotonicity properties as the function $\Fdefa(Y_1)$ in Figure \ref{fig:A_less_1}. We label the singular points as $Q_{\pm\pi}^{s,l}=(\pm\pi, \yl(\pi))$, $Q_{\pm\pi}^{c}=(\pm\pi,\ym(\pi))$, $Q_{\pm\pi}^{s,u}=(\pm\pi,\yu(\pi))$, where the $Y_1$ coordinates have the ordering  $\yl(\pi) < \ym(\pi) < \yu(\pi)$, and will use Morse theory in order to characterise the singular points.
       At $Q_0^s=(0,\ym(0))$,         \begin{align*} 
&D^2 H_1 (0,\ym(0)) =\begin{pmatrix}
&-M_1 g(\ym(0)) &0 \\
&0 & M_1 g(\ym(0)) -\gamma_1 
\end{pmatrix},
\end{align*}
with
$-M_1 g(\ym(0)) = M_1 \sqrt{1-A^2} \cosh \left( \ym(0) - \frac 12 \lAA \right)>0$, $M_1 g(\ym(0)) -\gamma_1 <0$, due to the monotonicity conferred by \eqref{g2der}. Hence $Q_0^s$ is indeed a saddle point. For the remaining singular points, we need to determine the nature of the eigenvalues of
         \begin{align*} 
&D^2 H_1 (\pi,Y_1 )=\begin{pmatrix}
&M_1 g(Y_1) &0 \\
&0 & -M_1 g(Y_1) -\gamma_1 
\end{pmatrix},
\end{align*}
for $Y_1$ being  $\yl(\pi), \ym(\pi)$, or $\yu(\pi)$. In order to do so, we will use the monotonicity properties of $F(\pi,\cdot)$ which match the one for $\Fdefa(Y_1)$, in Figure \ref{fig:A_less_1}.
Since $M_1 g (Y_1) <0, $ for any $Y_1$, due to  $A<1$ and $M_1>0,$ it suffices to examine the monotonicity of $F(X,\cdot)$ as it crosses the points $\yl(\pi), \ym(\pi)$ and $\yu(\pi)$, in order to conclude about the remaining critical points. Recalling Figure \ref{fig:A_less_1} and \eqref{Hessian_H1_fg}, it follows that $Q_{\pm\pi}^{s,l}=(\pm\pi, \yl(\pi))$ and $Q_{\pm\pi}^{s,u}=(\pm\pi,\yu(\pi))$ are saddle points, while $Q_{\pm\pi}^{c}= (\pm\pi,\ym(\pi))$ is a centre.  
\end{proof}
\begin{remark}
Physically relevant streamlines must lie between (and avoid) the singular points $Q_0^s$ and $Q_{\pm\pi}^{s,u}$, which provides additional physical conditions for the wave-steepness parameters $\e,\e_1$. Furthermore $\dot X<0$ along any physically relevant streamline in the $(X,Y_1)-$plane.
\end{remark}

   \begin{proposition}\label{P:Aless1-b}
   For $|A|<1$, let $\gamma_1<0$ (and so $\cdef>0$) and $|\mrl|<1$. If  \eqref{Aless1cond} holds and 
    \begin{equation}\label{Al1bcond}
F\left(X,\frac{1}{2}\lAA\right)= -{k\cdef}  -\frac{\gamma_1}{2} \lAA\neq 0,
\end{equation}
 then the dynamical system \eqref{Upper-DS}    has $5$ singular points.    \end{proposition}
   \begin{figure}[H]
\begin{center}
\resizebox{.7\textwidth}{!}{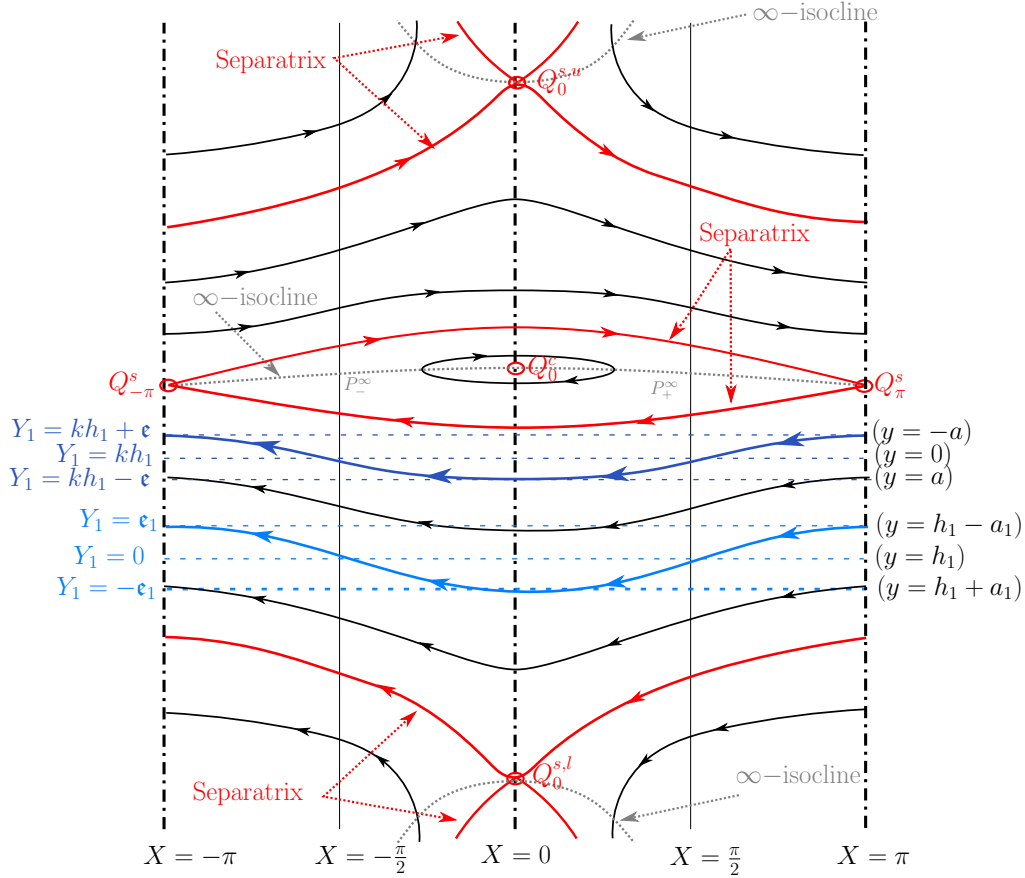}
\caption{Phase portrait for $A<1$, $\gamma_1<0$ ($\cdef>0$), where \eqref{Aless1cond} holds. The schematic corresponds to a regime whereby $\ym(X) > \frac 12 \lAA$ for all $X$. A similar legend applies as in Figure \ref{fig:PP-Aless1-a}.}
\label{fig:PP-Aless1-b}
\end{center}
\end{figure}
   \begin{proof}
The proof of Proposition \ref{P:Aless1-b} follows along the lines of Proposition \ref{P:Aless1-a}, with some essential differences that we outline briefly. First, $F(X,\cdot)$ shares similar monotonicity properties as $-\Fdefa (Y_1)$ of Figure \ref{fig:A_less_1} when $\cos X>0$. Since \eqref{Aless1cond} holds, we infer that the $\infty-$isocline comprises the graph of a function $\ym(X)$ which is well-defined for all $X$.  Using the fact that $ F(X,\ym(X)) = 0$,  and performing an implicit differentiation, we obtain:
\[
    \frac{\partial \ym}{\partial X}{\frac{\partial{ {F} }}{\partial Y_1}}(X,  \ym (X)) = {M_1 \sin Xf(\ym (X))},
  \]
for all $X$ in our domain, so  that the location of the line $Y_1=\frac 12 \lAA$ plays a key role in determining the monotonicity of $\ym$. As the monotonicity behaviour of $F(X_0,\cdot)$ matches that  of  $-\Fdefa (Y_1)$ when $\cos X_0 >0$, and $F(X_0,\cdot)$ is increasing otherwise, it follows that ${\frac{\partial{ {F} }}{\partial Y_1}}(X_0,  \ym (X_0)) > 0$, for all $X_0.$ Then,  if $\ym(X) < \frac 12 \lAA$ for all $X\in [-\pi,\pi]$, it follows from examining the sign of $f$ that $\ym$ is increasing for $X\in(0,\pi)$, and decreasing otherwise. In the case where the relative positions of the horizontal line $Y_1 = \frac 12 \lAA$ and the graph of $\ym$ are reversed, a reversal in the monotonicity properties of $\ym$ also takes place.  We note that the graph of $\ym$ does not intersect the horizontal line $Y_1 =\frac 12 \lAA$, since $F\left(X,\frac 12\lAA\right) = - k\cdef-  \frac{\gamma_1}{2} \lAA\neq 0$ by  \eqref{Al1bcond}. If \eqref{Al1bcond} does not hold then it follows that $\ym(X)$ is the horizontal line $Y_1=\frac{1}{2}\lAA$.
\end{proof}
  \begin{remark}
Physically relevant streamlines must lie between (and avoid) the singular points $Q_0^{s,l}$ and $Q_{\pm\pi}^{s}$, which provides additional physical conditions for the wave-steepness parameters $\e,\e_1$. Furthermore $\dot X<0$ along any physically relevant streamline in the $(X,Y_1)-$plane.
\end{remark}
  \begin{remark}
 For $|A|<1$ and $\cdef>0$, the phase portraits share the same essential features as those in the irrotational setting (cf. \cite{HV-JDE,HV-AN}) whenever  $|\mrl|>1$ and \eqref{Aless1cond} holds. A similar observation holds if  $|\mrl|<1$ and 
 \[
 \arcosh \frac{1}{|\mrl|}-\tanh \arcosh \frac{1}{|\mrl|} < \left| \frac{k\cdef}{\gamma_1} +\frac 12 \lAA\right|.
 \] In particular, for $\gamma_1>0$ and $\cdef>0$, a transition from 7 to 3 singular points occurs whenever the above relation, and so also \eqref{Aless1cond}, become equalities.
 \end{remark} 
 
\subsubsection{$|A|<1$, and $\cdef<0$ (critical layer present)}
\begin{proposition}\label{P:Aless1-c}
For $|A|<1$, let $\gamma_1>0$ be such that $\cdef<0$. Suppose $ |\mrl|<1$, \eqref{Aless1cond} holds, and also 
\[
k\cdef\neq -\frac{\gamma_1}{2}\lAA.
\] Then the $\infty-$isocline for the dynamical system  \eqref{Upper-DS} consists of $3$ disjoint parts, and there exist $5$ singular points.
\end{proposition}
  \begin{figure}[H]
\begin{center}
\resizebox{.7\textwidth}{!}{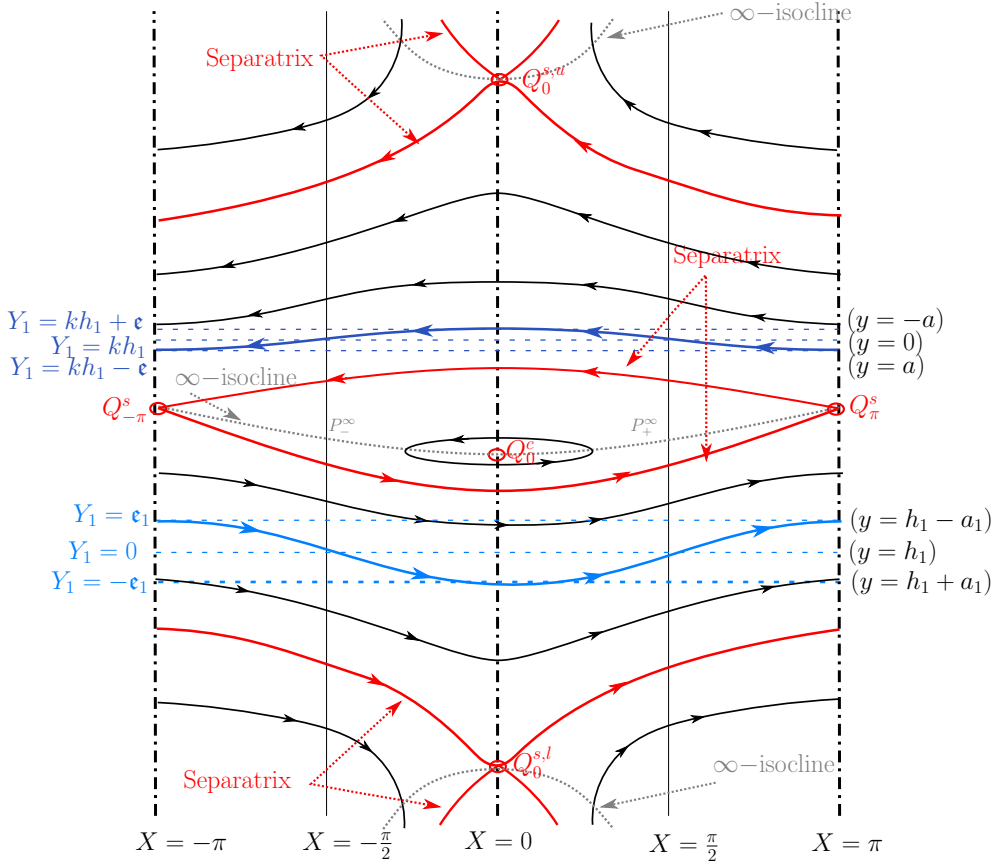}
\caption{Phase portrait for $A<1$, $\cdef<0$ and $\gamma_1>0$. The schematic corresponds to a regime whereby $\frac{k\cdef}{\gamma_1}+\frac 12\lAA<0$. A similar legend applies as in Figure \ref{fig:PP-Aless1-a}.}
\label{fig:PP-Aless1-c}
\end{center}
\end{figure}

\begin{proof}
  First, note that the line $Y_1=\frac 12 \lAA$ cannot intersect the $\infty-$isocline: for fixed $X$, we have $F(X,\frac 12 \lAA)=-k\cdef-\frac{\gamma_1}{2}\lAA,$ which is non-zero by assumption. Similarly,  the line $Y_1=0$ also cannot intersect the $\infty-$isocline: $F(X,0)=M_1 A \cos X- k\cdef$ and $M_1=\e_1 k\cdef \ll k\cdef$. In the regime $|A|<1,$ we have $\frac {\partial F}{\partial Y_1}<0$ for  $X\in[-\pi,\pi]\setminus(-\frac{\pi}{2},\frac{\pi}{2})$ by our assumptions on $A$ and $\gamma_1$.
On the other hand,  \eqref{Aless1cond} implies that $F(X,Y_1^+(X)) F(X,Y_1^-(X))<0$ for all $X$ in {$(-\frac{\pi}{2},\frac{\pi}{2})$},  and so  $F(X,\cdot)$  behaves similarly to $\Fdefa(Y_1)$. Consequently,   the equation $F(X,Y_1)=0$ has three solutions for any fixed $X\in \left(-\frac{\pi}{2},\frac{\pi}{2}\right)$, and one solution otherwise.
 
Let $\yl(X)<\ym(X)<\yu(X)$ be the three solutions of $F(X,Y_1)=0$, when $-\frac{\pi}{2}< X<\frac{\pi}{2}$ is fixed. Then,  suppressing any $X$ dependence,   the relations $\yl<Y_1^-<\ym$ and $\ym<Y_1^+<\yu$ hold true, as  $F$ does not vanish along $Y_1^{\pm}$, by  \eqref{Y1pm} and \eqref{Aless1cond}. In particular, we have 
\[
{\yl(X)< Y_1^{-}(X) < \frac 12 \lAA< Y_1^{+} (X) } < \yu(X) \quad \text{for all} \quad |X|<\frac{\pi}{2},
\] after taking into account \eqref{Y1pm}.
If  $Y_1(X)$ satisfies $F(X,Y_1(X))=0$, then an implicit differentiation yields
\begin{align}\label{Y1-Impl}
 \frac{d}{dX}F(X,Y_1(X))=0 \Leftrightarrow M_1 \sin X f\left(Y_1 (X)\right)  =\frac{\partial Y_1}{\partial X}\frac{\partial F}{\partial Y_1},
\end{align}
for all $X_0$, where $\frac{\partial F}{\partial Y_1}= M_1 \cos X g\left(Y_1  (X)\right)-\gamma_1$. Thus,  taking into account  $\frac 12 \lAA < Y_1^+ (X)< \yu (X)$,  Figure \ref{fig:A_less_1}, and examining \eqref{Y1-Impl}, we deduce that $\yl$ is a decreasing function for $0\leq X<\frac{\pi}{2}$, whereas $\yu$ is increasing in the same region. The monotonicity of both functions with respect to $X$ is reversed in the region $-\frac{\pi}{2}<X<0$.  Inspecting the limit of $F(X,\ym(X))$, as $X\to {\frac{\pi}{2}}^-$, we have $\lim_{X\to {\frac{\pi}{2}}^-} \ym(X)=-\frac{k\cdef}{\gamma_1}$.

 Let $\ym$ be the unique solution of the equation $F(X,Y_1)=0$, for fixed $X$ such that $\frac{\pi}{2}\leq|X|$. Revisiting our earlier arguments we deduce that $\lim_{X\to {\frac{\pi}{2}}^+} \ym(X)=-\frac{k\cdef}{\gamma_1}$. We now examine the relative position of the horizontal line $Y_1=\frac 12 \lAA$ and the graph of $\ym$. The monotonicity of $F(X,\cdot)$ dictates that, if 
 \[
0 = F(X, \ym(X))< F \left( X,\frac 12 \lAA \right) = -k\cdef- \frac{\gamma_1}{2} \lAA
\]
holds, then {$\ym(X)<\frac 12 \lAA$}, for all $X \in [-\pi,\pi]$. Furthermore,  $\ym$ is increasing for positive $X$, and decreasing for negative $X$, due to \eqref{Y1-Impl}. Analogously, when $\frac{k\cdef}{\gamma_1}+\frac 12\lAA>0$, the graph of $\ym(X)$ must lie above the line $Y_1=\frac 12 \lAA$ for all $X \in [-\pi,\pi]$, with a reversal in the monotonicity of $\ym$.

The singular points for the dynamical system \eqref{Upper-DS} occur at $Q_0^c=(0,\ym(0))$, $Q_0^{s,l}=(0,\yl(0))$, $Q_0^{s,u}=(0,\yu(0))$, $Q_{\pm\pi}^s=(\pm \pi,\ym(\pm \pi))$. Since $\frac{\partial {F}}{\partial Y_1}(0,\ym(0))=M_1  (A\sinh \ym(0) - \cosh \ym(0)) -\gamma_1$, the Hessian \eqref{Hessian_H1_fg} at $Q_0^c$ is 
\begin{align*}
D^2 H_1 (0,\ym(0))=\begin{pmatrix}
&-M_1 (A\sinh \ym(0) - \cosh \ym(0)) &0\\
&0&\frac{\partial {F}}{\partial Y_1}(0,\ym(0))
\end{pmatrix}.
\end{align*}
The first diagonal term is negative, since $M_1<0$ and $A<1$,  and the second diagonal term is also negative due to monotonicity considerations: hence Morse theory implies that $Q_0^c=(0,\ym(0))$ is a centre. Monotonicity considerations (cf. Figure \ref{fig:A_less_1}) imply $\frac{\partial {F}}{\partial Y_1}(0,\yl(0)),\frac{\partial {F}}{\partial Y_1}(0,\yu(0))>0$,  hence similar arguments imply that  $Q_0^{s,l}=(0,\yl (0))$ and $Q_0^{s,u}=(0,\yu(0))$ are saddle points. Finally, 
\begin{align*}
&D^2 H_1 (\pi,\ym( \pi))=\begin{pmatrix}
&M_1 g\left(\ym(\pi)\right) &0\\
&0&-g\left(\ym(\pi)\right)-\gamma_1
\end{pmatrix},
\end{align*}
with the Hessian matrix having the same form when $X=-\pi$. Noting that  $M_1<0$, $A<1$, and $\frac{\partial {F}}{\partial Y_1}(\pi,\ym(\pi))=-g\left(\ym(\pi)\right)-\gamma_1$, it follows that $Q_{\pm \pi}^{s}=(\pm \pi , \ym(\pm \pi))$ correspond to saddle points. 
\end{proof}
\begin{remark}
In the presence of the critical layer, and bearing in mind the monotonicity of $ {F}( X,\cdot)$, the existence of physically admissible solutions of \eqref{Upper-DS} corresponding to streamlines of the upper-fluid layer is equivalent to the conditions
 $ {F}(0,  \yl (0))<{F(0,- \e_1)}$ and  ${F(0,kh_1 -  \e)}  <{F} (0, \yu (0) )$ holding.
  \end{remark}


\subsection{The case $A>1$}\label{PP-A>1}
For $A>1$ we have
\begin{equation}\label{Cond3}
\frac{a}{a_1}\frac{c}{\cdef}<e^{-k h_1},
\end{equation}
since the right-hand side of \eqref{A-alt} is a decreasing function of $A$.
For $\cdef>0$, we may equivalently express \eqref{Cond3} as
\[
 \gamma_1< \frac{c}{h_1}\left(1-\frac{a}{a_1}e^{k h_1}\right),
\]
with the reverse inequality holding for $\cdef<0$. These relations are useful for immediately inferring various basic properties of the wave solutions. For instance, 
if $\gamma_1>0$ is such that $\cdef>0$, then we can have either in-phase waves with the restriction $a<a_1\cdef e^{-kh_1}/c$ (noting that $\cdef/c<1$ in this regime), or else out-of-phase waves.
Regarding the existence of out-of-phase waves in this regime (that is, in the absence of critical layers)
for $A>1$ it follows from direct calculation that 
\[
g( Y_1 )=  A \sinh Y_1 - \cosh Y_1=0 \iff Y_1=\frac{1}{2} \ln\Adefb,
\] where we denote $\Adefb:=\frac{A+1}{A-1}$. Hence, when $A>1$, the horizontal line $Y_1=\frac{1}{2} \ln\Adefb$ demarcates a change in direction in the vertical velocity for the dynamical system \eqref{Upper-DS}. Consequently, even in the absence of critical layers (and associated flow reversal) out-of-phase waves (whereby $a<0$) will occur if $\lA<kh_1$: this is in contrast to the previously considered situations when $A\leq 1$.
The phase-plane analysis we perform below furnishes us with an elegant and explicit visualisation of the transition process from in-phase, to out-of-phase, internal--surface wave coupling.

\begin{figure}[H]
\begin{center}
\resizebox{.42\textwidth}{!}{
\begingroup%
  \makeatletter%
  \providecommand\color[2][]{%
    \errmessage{(Inkscape) Color is used for the text in Inkscape, but the package 'color.sty' is not loaded}%
    \renewcommand\color[2][]{}%
  }%
  \providecommand\transparent[1]{%
    \errmessage{(Inkscape) Transparency is used (non-zero) for the text in Inkscape, but the package 'transparent.sty' is not loaded}%
    \renewcommand\transparent[1]{}%
  }%
  \providecommand\rotatebox[2]{#2}%
  \newcommand*\fsize{\dimexpr\f@size pt\relax}%
  \newcommand*\lineheight[1]{\fontsize{\fsize}{#1\fsize}\selectfont}%
  \ifx\svgwidth\undefined%
    \setlength{\unitlength}{257.36301933bp}%
    \ifx\svgscale\undefined%
      \relax%
    \else%
      \setlength{\unitlength}{\unitlength * \real{\svgscale}}%
    \fi%
  \else%
    \setlength{\unitlength}{\svgwidth}%
  \fi%
  \global\let\svgwidth\undefined%
  \global\let\svgscale\undefined%
  \makeatother%
  \begin{picture}(1,0.48989612)%
    \lineheight{1}%
    \setlength\tabcolsep{0pt}%
    \put(0.88215606,0.38149611){\makebox(0,0)[lt]{\lineheight{0}\smash{\begin{tabular}[t]{l}\tb{\Large{$\Fdefb(Y_1)$}}\end{tabular}}}}%
    \put(1.45297257,-0.70694931){\color[rgb]{0,0,0}\makebox(0,0)[lt]{\lineheight{1.25}\smash{\begin{tabular}[t]{l}x\end{tabular}}}}%
    \put(0,0){\includegraphics[width=\unitlength,page=1]{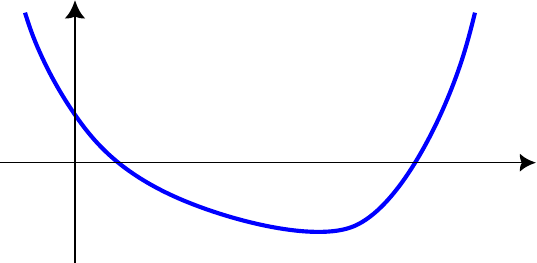}}%
    \put(1.0148198,0.14791532){\makebox(0,0)[lt]{\lineheight{0}\smash{\begin{tabular}[t]{l}\Large{$Y_1$}\end{tabular}}}}%
    \put(0,0){\includegraphics[width=\unitlength,page=2]{Fbmono.pdf}}%
    \put(0.60247112,-0.00789005){\makebox(0,0)[lt]{\lineheight{1.25}\smash{\begin{tabular}[t]{l}\Large{${Y_1^+}$}\end{tabular}}}}%
  \end{picture}%
\endgroup%
}
\caption{The function $\Fdefb  (Y_1)$, when $A>1$.}
\label{fig:A_ge_1}
\end{center}
\end{figure}

The $0-$isocline of the dynamical system \eqref{Upper-DS}, which consists of the points for which $\dot Y_1=G(X,Y_1)=M_1\sin Xg(Y_1)=0$,
comprises the vertical lines $X=0,\pm \pi$, and the horizontal line $Y_1=\frac{1}{2} \ln\Adefb$. 
The $\infty-$isocline of the dynamical system \eqref{Upper-DS}  consists of points for which $\dot X=F(X,Y_1)=M_1\cos X f(Y_1)-k\cdef -\gamma_1  Y_1=0$, where $f(Y_1)=  A \cosh Y_1 - \sinh Y_1$. As before, the points $P_{\pm}^{\infty}=\left(\pm \frac{\pi}{2},-\frac{k\cdef}{\gamma_1}\right)$ in \eqref{PPM} will be in the $\infty-$isocline. Qualitative features of the $\infty-$isocline can be determined by considering, for fixed $X$, the monotonicity properties of $F(X,\cdot)$ via
$\frac{\partial F}{\partial Y_1}=M_1\cos X g(Y_1)-\gamma_1$. Since $g(Y_1)$ is a strictly increasing function of $Y_1$ for $A>1$, it follows that if $M_1\cos X>0$, then $F(X,\cdot)$ is  strictly decreasing until it reaches its minimum value $Y_1^+$, from which point it is strictly increasing: $F(X,\cdot)$ has monotonicity properties similar to  $\Fdefb (Y_1)$ in Figure \ref{fig:A_ge_1}; if $M_1\cos X<0$, the reverse scenario occurs, with $F(X,\cdot)$ having similar monotonicity properties as  $-\Fdefb (Y_1)$ in Figure \ref{fig:A_ge_1}. 
Hence, for a given $X$, the $\infty-$isocline contains points with abscissa $X$ if $F(X,\cdot)$  intersects the $Y_1-$axis, which is equivalent to $F (X,Y_1^+)$ either touching the $Y_1-$axis, or else lying on the opposite side to the $Y_1-$axis as the asymptotic values of $F(X,\cdot)$: the latter scenario is depicted in Figure \ref{fig:A_ge_1}.
The extremum value $Y_1^+$ for $F(X,\cdot)$ can be computed (for $X\neq \frac{\pi}{2}$) from
\begin{align*}
\frac{\partial {F} }{\partial Y_1}(X,Y_1^+(X))=0
\iff
g(Y_1^+)=A \sinh Y_1^+ - \cosh Y_1^+=\frac{\gamma_1}{M_1\cos X} 
\\
\iff (A-1)\left(e^{Y_1^+}\right)^2-\frac{2\gamma_1}{M_1\cos X} e^{Y_1^+}-(A+1)=0
\\
\iff e^{Y_1^+}=\sqrt{\Adefb}\left(s+\sqrt{s^2+1}\right), \quad s=\frac{\gamma_1}{M_1\cos X\sqrt{A^2-1}}. 
\end{align*}
Using the relation $\arsinh(s)=\ln \left(s+{\sqrt {s^{2}+1}}\right)$ we get, for $X \neq \pm \frac{\pi}{2}$,
\begin{equation}\label{Y1+}
Y_1^{+}(X)=\frac 1 2 \ln\Adefb+ \arsinh z_2(X),  \mbox{ for } z_2(X):=\frac{\gamma_1}{M_1 \cos X\sqrt{A^2-1}}.
\end{equation}
  Note that since $g(Y_1)$ is an increasing function it follows that
(for  $X\neq \pm \frac{\pi}{2}$)
 \begin{align*}
Y_1^+(X)<0\iff\sqrt{\Adefb}\left(z_2(X)+\sqrt{z_2^2(X)+1}\right)<1
\\
\iff \frac{\gamma_1}{M_1 \cos X}=g(Y_1^+)<g(0)=-1,
\end{align*}
where we note that the second relation implies that  $z_2(X)<0$. Since \eqref{A-alt} implies that $\frac{a}{a_1} \frac{c}{\cdef}=-g(kh_1)$, it follows that
\begin{align*}
Y_1^+(X)>kh_1 \iff \frac{\gamma_1}{M_1 \cos X}=g(Y^+_1)>g(kh_1)=-\frac{a}{a_1} \frac{c}{\cdef}.
\end{align*}
Furthermore,  $Y_1^+$ never intersects the $0-$isocline given by the line $Y_1=\lA$ (as $z_2(X)\neq 0$), and  $\lim_{X\to \pm \frac{\pi}{2}}\left|Y_1^+(X)\right|=\infty$.  Using hyperbolic function identities, and the fact that $f(\lA)=\sqrt{A^2-1}$ and $g(\lA)=0$, we get
  \begin{align}\label{g2A>1}
 	& F (X,Y_1^+(X))=M_1\cos X f(\Adefb) \cosh \arsinh z_2(X)- k \cdef - \gamma_1  Y_1^+ \nonumber\\
& =  M_1 \cos X  \sqrt{A^2 -1}\sqrt{z_2 (X)^2 + 1} - k \cdef - \gamma_1  Y_1^+ \nonumber\\
&=\gamma_1\left(\frac{\sqrt{z_2 (X)^2 + 1} } {z_2 (X) }  -\arsinh z_2(X) - \frac{k\cdef}{\gamma_1}- \lA\right) \nonumber \\
&=\gamma_1\left(\coth \arsinh z_2(X) -\arsinh z_2(X) - \frac{k\cdef}{\gamma_1}-  \lA \right),
 \end{align} 
using the identities $\cosh \arsinh s = \sqrt{s^2+1}$ and  $\coth \arsinh s = \frac{\sqrt{s^2+1}}{s}$ ($s\neq 0$). The function $s\mapsto \coth s -s$ is such that  $\lim_{s\to\pm \infty}\coth s - s = \mp \infty$, therefore, for values of  $X$ close to $\frac{\pi}{2}$, $F (X,Y_1^+)$ must take the opposite sign to the asymptotic values of  ${F} (X,Y_1)$, implying that the equation ${F} (X,Y_1)=0$ has two solutions in $Y_1$ (and so the $\infty-$isocline has two branches in this region).
Finally, the definition of $z_2$ in \eqref{Y1+}, together with \eqref{g2A>1}, imply that
\begin{equation}\label{g2A>1_prop}
{F} (X+\pi,Y_1^+(X+\pi))=-{F} (X,Y_1^+(X))-2\left( k\cdef +  \frac {\gamma_1} 2 \ln\Adefb\right)\!, \ X\neq \pm \frac{\pi}{2}.
\end{equation}

 \subsubsection{$A>1$ and $\cdef>0$ (no critical layers)}
\begin{proposition}\label{P:Agreat1-a} 
    For $A>1$,  let $\gamma_1>0$ be such that $\cdef>0$. If
           \begin{equation}\label{condA1}
          F\left(\pi,Y_1^+ (\pi)\right)>0,  
       \end{equation}
       and
        \begin{equation}\label{condA1b}
 F(0,-\e_1)<0
       \end{equation}
       are satisfied, then the dynamical system \eqref{Upper-DS}   
 has $6$ singular points.        \end{proposition}
  \begin{figure}[H]
\begin{center}
\resizebox{.665\textwidth}{!}{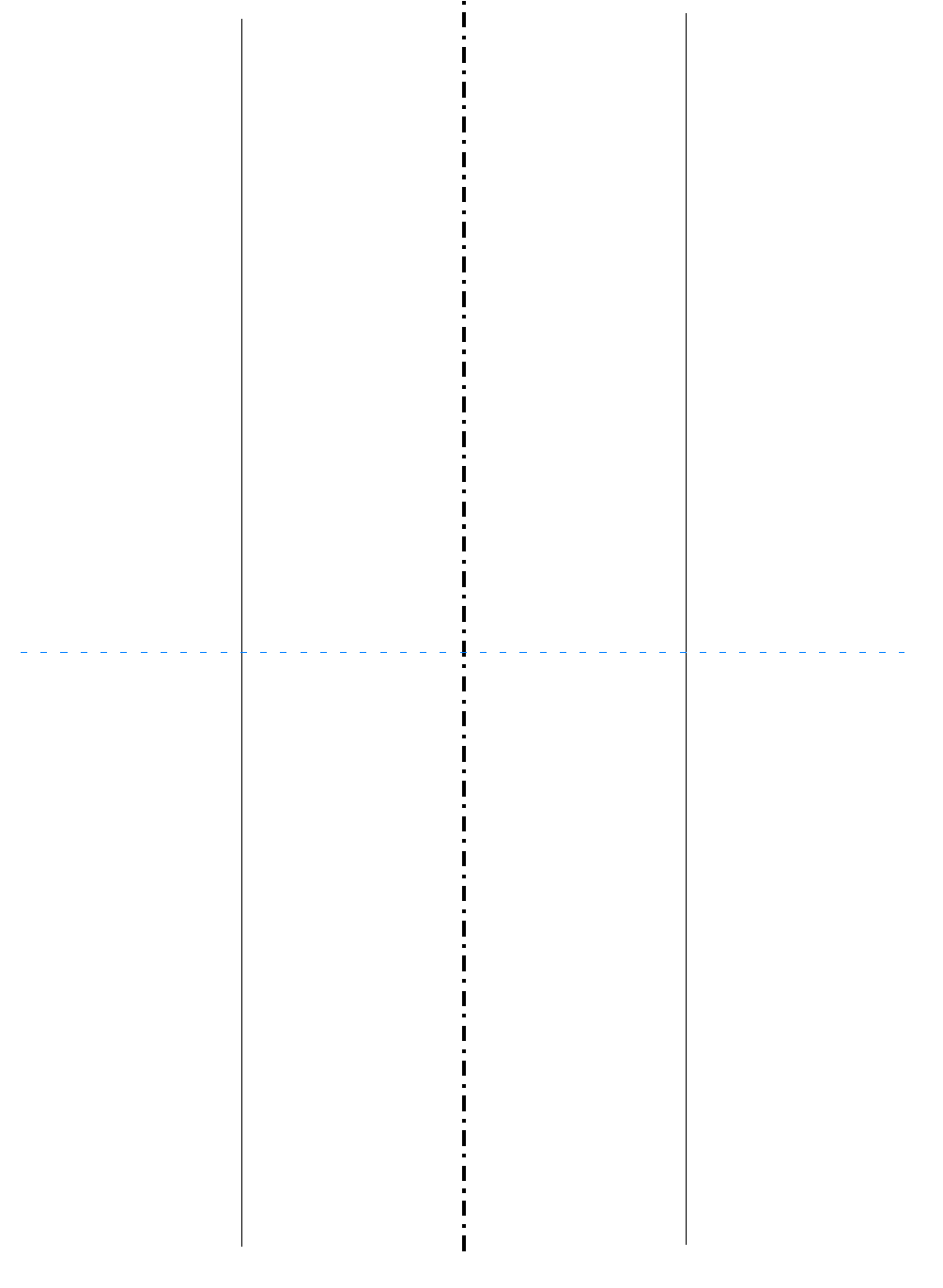}
\caption[C2]{Phase portrait for $A>1$, $\cdef>0$, $\gamma_1>0$, where \eqref{condA1} and \eqref{condA1b} hold. The dotted grey lines \tikz[baseline=-0.5ex] \draw[color=gray, very thick, densely dotted] (.0,0) -- (.4,0) ; represent $\infty-$isoclines, with the dotted-dashed lines \tikz[baseline=-0.5ex] \draw[thick,dash dot] (.0,0) -- (.5,0) ;  representing $0-$isoclines. The {surface wave} profile (\tikz[baseline=-0.5ex] \draw[color=azure, very thick] (.0,0) -- (.7,0);) has mean-water level $Y_1=0$, corresponding to $y=h_1$. The internal wave profile is illustrated for two differing values of the mean water level $Y=kh_1$, corresponding to $y=0$: $kh_1 \leq \bar{Y_1}$ in Case 1 (\tikz[baseline=-0.5ex] \draw[color=bleudefrance, very thick] (.0,0) -- (.7,0);), whereas  {$kh_1 > \bar{Y_1}$} for Case  2 (\tikz[baseline=-0.5ex] \draw[color=ballblue, very thick] (.0,0) -- (.7,0);).}
\label{fig:PP-Agreat1-a}
\end{center}
\end{figure}
\begin{proof}
As discussed in the previous Section,  since $M_1=\e_1k\cdef>0$  the behaviour  of $ F ( X, \cdot ) $ is similar to that of $\Fdefb$ depicted in Figure \ref{fig:A_ge_1} when $|X| < \frac {\pi} 2$, and similar to that of   $ -\Fdefb$  when $|X| \in\left(\frac {\pi} 2,\pi\right]$. For $X$ in the latter region (with $\cos X <0$), we claim that $ F (X,Y_1^+(X)) > 0$ holds, where $\frac{\partial F }{\partial Y_1}(X,Y_1^+(X))=0$. This follows from noting that \eqref{condA1} can be reexpressed 
  \begin{equation}\label{condA1a}
\arsinh \frac{1}{\tilde{\mathfrak{R}}}-\coth \arsinh \frac{1}{\tilde{ \mathfrak{R}}} >  \frac{k\cdef}{\gamma_1} +\lA, \ \mbox{ for } \tilde{\mathfrak{R}}: = \frac{M_1 \sqrt{A^2-1} }{\gamma_1},
 \end{equation}
 coupled with the observation that the function $ X \mapsto \coth  \arsinh \frac 1 {\tilde{\mathfrak R}\cos X} -  \arsinh \frac 1 {\tilde{\mathfrak R}\cos X} $ restricted to the interval $[-\pi,-\frac \pi 2)$ attains its minimum at $x_0=-\pi$. Consequently,  for fixed $X$ such that $|X| \in\left(\frac {\pi} 2,\pi\right]$, the equation $F ( X, Y_1 ) =0$ has two solutions with respect to $Y_1$. 
We can then apply \eqref{g2A>1_prop} directly to infer that the reverse scenario, $ F (X,Y_1^+(X)) < 0$, holds  for $X$ in the region $ | X | < \frac { \pi } 2$. This proves that, for any fixed $ X \neq \pm \frac { \pi } 2$, the equation $ F ( X, Y_1 ) = 0 $ has two solutions with respect to the  $Y_1$ variable. Since $\frac{\partial{{F} }}{\partial Y_1}$ vanishes only along $Y_1^+$ and ${F} (X,Y_1^+(X))\neq 0$, the implicit function theorem implies that we can write the solutions  as functions of $X$.
For $|X| \in\left(\frac {\pi} 2,\pi\right]$ we denote   these solutions by  $\yl(X) < Y_1^+ (X) < \ym(X)$, whereas for  $|X|< \frac { \pi } 2$ we denote the solutions as $\ym(X)< Y_1^+ (X) < \yu(X)$.

If  we define $\ym\left(\pm \frac \pi 2\right)=-\frac{k\cdef}{\gamma_1}$, then $\ym(X)$ defines a curve which extends across the entire domain $[-\pi,\pi]$. The monotonicity of $\ym$ can be established from its slope which is prescribed by
\begin{align*}
 \frac{d}{dX}F(X,\ym(X))=0 \Leftrightarrow M_1 \sin X f( \ym (X) )=\frac{\partial \ym }{\partial X} \left(  M_1 \cos Xg\!\left(\ym(X)\right)-\gamma_1\right)\!,
\end{align*}
which allows us to deduce that $\ym$ is decreasing when $X_0\in (0,\pi)$, and increasing when $X_0\in (-\pi,0)$.  Since $-\frac{k\cdef}{\gamma_1}<0$, this implies that $\ym(X)$ must lie beneath the upper fluid domain, in terms of $(X,Y_1)$ coordinates,  for $|X|\leq \frac{\pi}{2}$. Furthermore, \eqref{condA1b} implies that $\ym(0)<-\e_1$, and the monotonicity properties of the free-surface streamline, allied to those of $\ym(X)$, imply that  $\ym(X)$ also lies beneath the upper fluid domain, in terms of $(X,Y_1)$ coordinates,  for $|X|\in \left(\frac {\pi} 2,\pi\right]$.

Bearing in mind the form of the $0-$isocline, there exist singular points at $Q_{\pm \pi}^{s}=(\pm\pi,\yl(\pm\pi))$ and $Q_{\pm \pi}^{c}=(\pm\pi,\ym(\pm\pi))$ (recalling that  $\yl(\pi) < Y_1^+(\pi) <\ym(\pi)$) as well as  at  $Q_{0}^{s,l}=(0,\ym(0))$ and $Q_{0}^{s,u}=(0,\yu(0))$ (where $\ym(0) < Y_1^+ (0) <  \yu(0)$).
There are no other singular points since the $\infty-$isocline does not intersect the horizontal line $Y_1=\lA$: we have already established that  $\ym(X)<0<\lA$ in the entire domain, and  $\yl(X) < Y_1^+ (X) < \ym(X)$ in  $|X| \in\left(\frac {\pi} 2,\pi\right]$; while in the region  $|X|\leq \frac{\pi}{2}$ we have $\yu(X) > Y_1^+ (X)>\lA$, where the latter inequality follows from \eqref{Y1+}.

 We examine the Hessian of the Hamiltonian $H_1$ at the singular points. For  $Q_0^{s,l}=(0,\ym(0))$ we have
         \begin{align*} 
&D^2 H_1 (0,\ym(0)) =\begin{pmatrix}
&-M_1 g(\ym(0)) &0 \\
&0 & M_1 g(\ym(0)) -\gamma_1 
\end{pmatrix}.
\end{align*}
The first diagonal term is positive, as $M_1>0$ and $\ym(0) < \lA$, while previously outlined monotonicity considerations imply that the second term in the diagonal is negative, therefore Morse theory implies that $Q_0^{s,l}=(0,\ym(0))$ is a saddle point. For $Q_0^{s,u}=(0,{\yu(0)})$ we have
 \begin{align*} 
&D^2 H_1 (0,{\yu(0)}) =\begin{pmatrix}
&-M_1 g({\yu(0)} ) &0 \\
&0 & M_1 g( {\yu(0)} ) -\gamma_1 
\end{pmatrix},
\end{align*}
and since $\lA<Y_1^+(0)< {\yu(0)} $ we have $-M_1 g(\yu(0))<0$, with the monotonicity properties yielding the claim. 
 For $Q_{\pi}^{s}=(\pi,\yl(\pi))$ we have
         \begin{align*} 
&D^2 H_1 (\pi,\yl(\pi)) =\begin{pmatrix}
&M_1 g(\yl(\pi)) &0 \\
&0 & -M_1 g(\yl(\pi)) -\gamma_1 
\end{pmatrix},
\end{align*}
the first diagonal term is negative since $\yl(\pi) < 0< \lA$, while monotonicity properties imply that the second term in the main diagonal is positive: $Q_{\pi}^{s,u}=(\pi,\yl(\pi))$  corresponds to a saddle point.
For     $Q_\pi^{c}=(\pi,\ym(\pi))$,
     \begin{align*} 
&D^2 H_1 (\pi,\ym(\pi)) =\begin{pmatrix}
&M_1 g(\ym(\pi)) &0 \\
&0 & -M_1 g(\ym(\pi)) -\gamma_1 
\end{pmatrix}.
\end{align*}
 Taking into account the fact that $Y_1^+(\pi)<{\ym(\pi)} < \lA$, and using monotonicity properties once more, we can deduce from Morse theory that  $Q_\pi^{c}=(\pi,{\ym(\pi)})$ corresponds to a centre. 
We note that since the upper fluid domain lies between the separatrices which emanate from $Q_0^{s,l}$, and $Q_0^{s,u}$, respectively, we have $\dot X<0$ along physically relevant streamlines.
\end{proof}
\begin{remark}
The arguments of Proposition \ref{P:Agreat1-a} imply that $2$ saddle-centre bifurcations occur when 
\[
 F\left(\pi,Y_1^+ (\pi)\right) =0 \iff \arsinh \frac{1}{\tilde{\mathfrak{R}}}-\coth \arsinh \frac{1}{\tilde{\mathfrak{R}}}  = \frac{k\cdef}{\gamma_1} +\lA,
 \] with a resulting transition from $6$ to $2$ critical points.
\end{remark}

\begin{proposition}\label{P:Agreat1-b}
  For $A>1$,  let $\gamma_1<0$ (and so $\cdef>0$). Let $\mathcal M=\max\{kh_1-\e,\lA\}$. If
       \begin{equation}\label{condA2}
          F\left(\pi,Y_1^+ (\pi)\right) >0,
       \end{equation}
and
        \begin{equation}\label{condA2b}
F(0,\mathcal M)<0, \ F(0,-\e)<0,
       \end{equation}
      are satisfied, then the dynamical system \eqref{Upper-DS}   
 has $6$ singular points. 
    \end{proposition}
           \begin{figure}[H]
\begin{center}
\resizebox{.7\textwidth}{!}{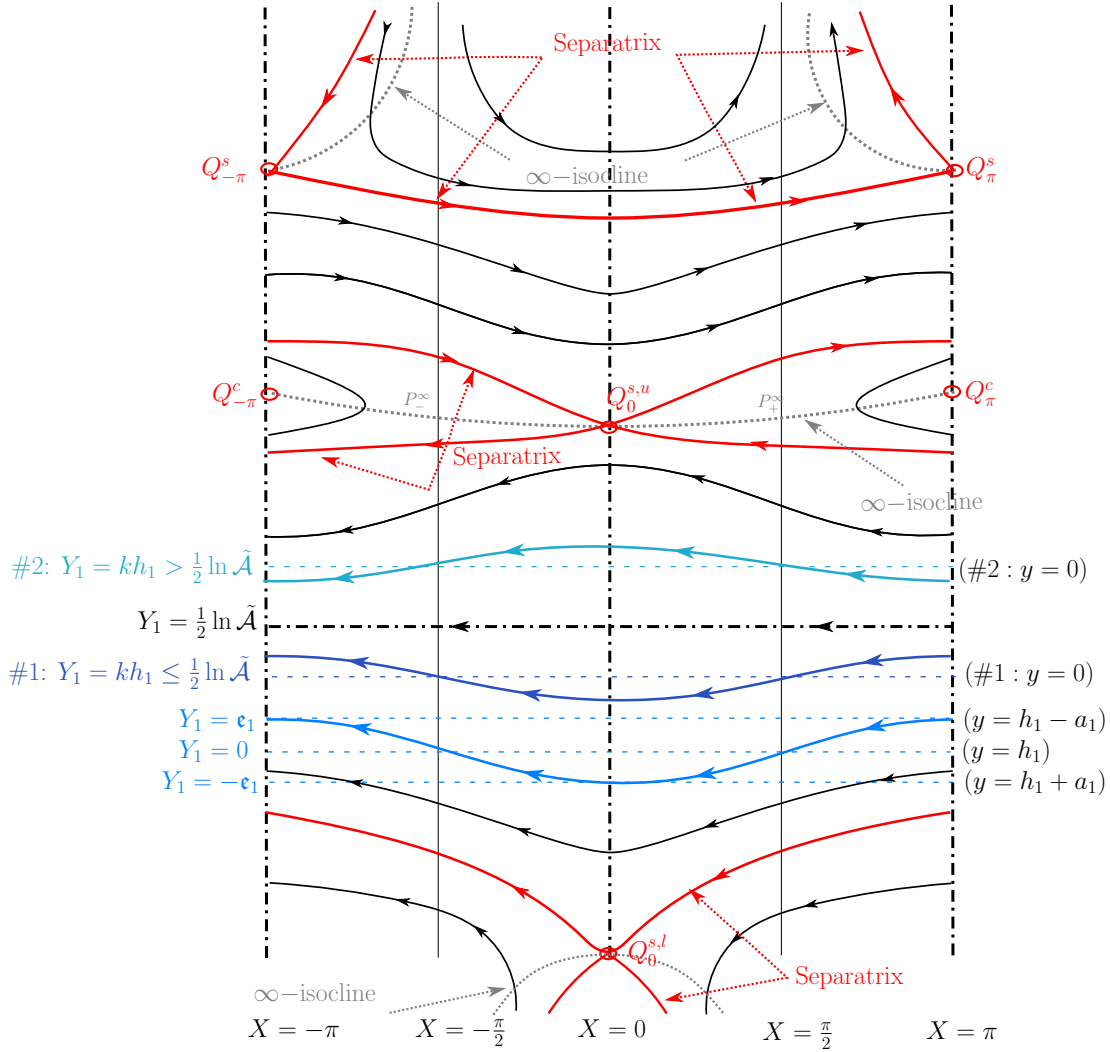}
\caption{Phase portrait for $A>1$, $\gamma_1<0$ ($\cdef>0$), where \eqref{condA2} and \eqref{condA2b} hold. A similar legend applies as in Figure \ref{fig:PP-Agreat1-a}.}
\label{fig:PP-Agreat1-b} 
\end{center}
\end{figure}
\begin{proof}
Condition \eqref{condA2} can be expressed as
 \[
 F\left(\pi,Y_1^+ (\pi)\right)>0\iff \arsinh \frac{1}{\tilde{\mathfrak{R}}}-\coth \arsinh \frac{1}{\tilde{\mathfrak{R}}} <  \frac{k\cdef}{\gamma_1} +\lA,
   \]
       and it is clear from the functional form of \eqref{g2A>1} that a similar inequality, $F(X,Y_1^+(X))>0$, will hold for all $|X|\in \left(\frac{\pi}{2},\pi\right]$ since $\tilde{\mathfrak R}<0$.
Condition \eqref{condA2b} implies that  $F\left(0,Y_1^+ (0)\right)<0$, which can be reexpressed as
\[
 F\left(0,Y_1^+ (0)\right)<0\iff
\coth \arsinh  \frac{1}{\tilde{\mathfrak{R}}} -\arsinh  \frac{1}{\tilde{\mathfrak{R}}} > \frac{k\cdef}{\gamma_1}+  \lA.
\]
From $\tilde{\mathfrak R}<0$,   and the functional form of \eqref{g2A>1}, it is clear that  $F(X,Y_1^+(X))<0$ will hold for all $|X|<\frac{\pi}{2}$. Therefore the equation $ F( X, Y_1 ) =0 $ has two solutions with respect to $Y_1$ for all $ X \neq \frac{\pi} 2 $,  and since $\frac{\partial{{F} }}{\partial Y_1}$ vanishes only along $Y_1^+$ and ${F} (X,Y_1^+(X))\neq 0$, the implicit function theorem implies that we can write the solutions  as functions of $X$. For $|X| \in\left(\frac {\pi} 2,\pi\right]$ we denote   these solutions by  $\ym(X) < Y_1^+ (X) < \yl(X)$, whereas for  $ | X | < \frac { \pi } 2$ we denote the solutions as $\yu(X)< Y_1^+ (X) < \ym(X)$ (note the reversal in ordering compared to Proposition \ref{P:Agreat1-a}).

We again set $\ym\left(\pm \frac \pi 2\right)=-\frac{k\cdef}{\gamma_1}$, in which case $\ym(X)$ defines a curve which extends across the entire domain $[-\pi,\pi]$. Similar to Proposition \ref{P:Agreat1-a}, we can establish monotonicity properties for  $\ym$, except now $\ym$ is increasing when $X_0\in (0,\pi)$, and decreasing for $X_0\in (-\pi,0)$. 
Since $-\frac{k\cdef}{\gamma_1}=-\frac{kc}{\gamma_1}+kh_1>kh_1>0$, as $\gamma_1<0$, it follows that $\ym(X)$ lies above the upper-fluid layer for $|X|\in \left[\frac{\pi}{2},\pi\right]$. Indeed, the first inequality in \eqref{condA2b} implies that $\ym(0)>kh_1-\e$, $\ym(0)>\lA$ (where we note that $kh_1-\e>\lA$ only when waves are out-of-phase, in which case $\e<0$) and the monotonicity properties of the internal-wave streamline, allied to those of $\ym(X)$, imply that $\ym(X)$ must lie above the upper fluid domain, in terms of $(X,Y_1)$ coordinates, for $|X|\leq \frac{\pi}{2}$. 

Due to the form of the $0-$isocline there exist singular points at $Q_0^{s,l}=(0, \yu(0))$, $Q_0^{s,u}=(0, \ym(0))$ (where $\yu(0) < Y_1^+(0)<\ym(0)$), and $Q_{\pm\pi}^{s}=(\pm\pi,\yl(\pm\pi))$, $Q_{\pm \pi}^c=(\pm \pi ,\ym(\pm \pi))$ (with $\ym(\pi) < Y_1^+(\pi)<\yl(\pi)$). There are no other singular points since we have shown that $\ym(X)$ lies above the horizontal line $Y_1=\lA$, with $\yl(X)>\ym(X)$ for $|X|\in \left[\frac{\pi}{2},\pi\right]$, while the second inequality in condition \eqref{condA2b} implies that $\yu(X)<0$: therefore the $\infty-$isocline cannot intersect the $0-$isocline along the line $Y_1=\lA$.

We examine the Hessian of the Hamiltonian $H_1$ at the singular points. At $Q_0^{s,u}=(0, \ym(0))$ we have
         \begin{align*} 
&D^2 H_1 (0,\ym(0)) =\begin{pmatrix}
&-M_1 g(\ym(0)) &0 \\
&0 & M_1 g(\ym(0)) -\gamma_1 
\end{pmatrix},
\end{align*}
where the first  diagonal term is negative, as $M_1>0$ and $\ym(X) > \lA $, while monotonicity considerations imply that the second diagonal term is positive: Morse theory gives that $Q_0^{s,u}=(0, \ym(0))$ is a saddle point. Similarly, at $Q_0^{s,l}=(0, \yu(0))$ we have
         \begin{align*} 
&D^2 H_1 (0,\yu(0)) =\begin{pmatrix}
&- M_1 g(\yu(0)) &0 \\
&0 & M_1 g(\yu(0)) -\gamma_1 
\end{pmatrix},
\end{align*}
and since $\yl(X) < 0< \lAA$ the first diagonal term is positive, and monotonicity properties imply that the second term in the main diagonal is negative: $Q_0^{s,l}=(0, \yu(0))$ corresponds to a saddle point. For  $Q_{\pi}^c=(\pi ,\ym(\pi))$,
 \begin{align*} 
&D^2 H_1 (\pi,\ym(\pi)) =\begin{pmatrix}
& M_1 g(\ym(\pi)) &0 \\
&0 & - M_1 g(\ym(\pi)) -\gamma_1 
\end{pmatrix},
\end{align*}
and the fact that $ \ym(\pi) > \lAA$, coupled with monotonicity properties, implies that  $(\pi,\ym(\pi))$ corresponds to a centre, due to Morse theory. Finally, at $Q_{\pi}^{s}=(\pi,\yl(\pi))$ we have
 \begin{align*} 
&D^2 H_1 (\pi,\yl(\pi)) =\begin{pmatrix}
& M_1 g(\yl(\pi)) &0 \\
&0 & - M_1 g(\yl(\pi)) -\gamma_1 
\end{pmatrix},
\end{align*}
and since $\lA< \yl(\pi)$ the first main diagonal term is positive, with the second diagonal term negative due to monotonicity properties: hence $Q_{\pi}^{s}=(\pi,\yl(\pi))$  is a saddle.
   \end{proof}
   \begin{remark}
It follows from the arguments of Proposition \ref{P:Agreat1-b} that  $2$ saddle-centre bifurcations occur when 
 \[
 F\left(\pi,Y_1^+ (\pi)\right)=0\iff \arsinh \frac{1}{\tilde{\mathfrak{R}}}-\coth \arsinh \frac{1}{\tilde{\mathfrak{R}}} =  \frac{k\cdef}{\gamma_1} +\lA,
   \]
in which case there is a transition from $6$ to $2$ critical points.
\end{remark}

\subsubsection{$A>1$, and $\cdef<0$ (critical layer present)}

\begin{proposition}\label{P:Agreat1-c}
For $A>1$, let $\gamma_1>0$ be such that $\cdef<0$.
 Suppose that the conditions
 \begin{equation}\label{A>1-in1}
 F \left(\pi,\lA\right)<0
 \end{equation}
 and
\begin{equation}\label{A>1-in}
F (0,Y_1^+(0))>0
\end{equation}      are satisfied. Then   the $ \infty-$isocline for the dynamical  system  \eqref{Upper-DS} consists of $4$ disjoint parts, and there exist $6$ singular points for \eqref{Upper-DS}, with $3$ of them forming a critical layer. 
\end{proposition}
     \begin{figure}[H]
\begin{center}
\resizebox{.7\textwidth}{!}{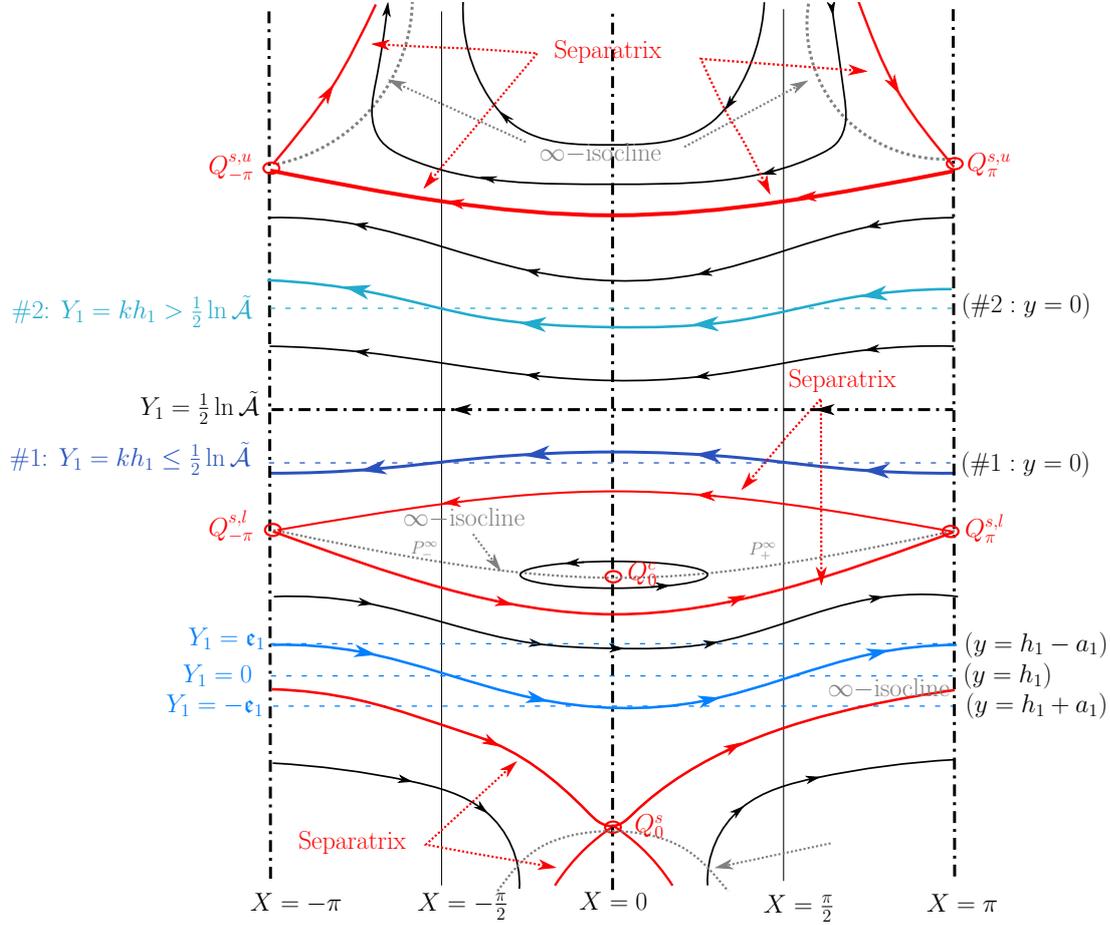}
\caption{Phase portrait for $A>1$, $\cdef<0$ and $\gamma_1>0$, where conditions \eqref{A>1-in1} and \eqref{A>1-in} hold. A similar legend applies as in Figure \ref{fig:PP-Agreat1-a}.}
\label{fig:PP-Agreat1-c}
\end{center}
\end{figure}
\begin{proof}
Since $\cdef<0$ we now have $M_1<0$  and, as the sign of $M_1\cos X$ determines the monotonicity properties of $F(X,\cdot)$, it follows that $F(X,\cdot)$ is qualitatively similar to the function $\Fdefb (Y_1)$ depicted in Figure \ref{fig:A_ge_1} for $|X|\in\left(\frac{\pi}{2},\pi\right]$, and qualitatively similar to $-\Fdefb (Y_1)$ for $|X|<\frac{\pi}{2}$.
We claim that condition  \eqref{A>1-in} implies that ${F} (X,Y_1^+(X))>0$ for $|X|<\frac {\pi} 2$.  Rewriting   \eqref{A>1-in} as
\[
\coth \arsinh z_2(0) - \arsinh z_2(0)> \frac{k\cdef}{\gamma_1}+  \frac 1 2 \ln\Adefb,
\]
we note that $z_2$ is decreasing when $X\in[0,\frac{\pi}{2})$ (since  $\gamma_1>0$, $M_1<0$) and so $\coth \arsinh z_2(\cdot) - \arsinh z_2(\cdot)$ is an increasing function in this range, thereby proving our claim. Bearing in mind the monotonicity properties of $F(X,\cdot)$ for fixed $|X|<\frac{\pi}{2}$, it follows that there are two solutions of $F(X,Y_1)=0$ with respect to the $Y_1$ variable for $X$ in this range.
 Condition  \eqref{A>1-in1} implies that 
 \begin{equation}\label{A>1-in1-in} 
k\cdef + \frac{\gamma_1}{2} \ln \Adefb>-M_1\sqrt{A^2-1}>0,
\end{equation}
and applying \eqref{g2A>1_prop} now gives us that ${F} (X,Y_1^+(X))<0$ for all $|X|\in(\frac{\pi}{2},\pi]$. This fact, coupled with the monotonicity properties of $F(X,\cdot)$, imply that there are two solutions of $F(X,Y_1)=0$ with respect to the $Y_1$ variable for fixed $X$ in this range.
Since $\frac{\partial{{F} }}{\partial Y_1}$ vanishes only along $Y_1^+$ and ${F} (X,Y_1^+(X))\neq 0$, the implicit function theorem implies that we can write these solutions as functions of $X$.

 Let us denote the branches of the $\infty-$isocline by $\yl(X)<Y_1^+(X)<\ym(X)$, when $|X|<\frac{\pi}{2}$, and $\ym(X) <Y_1^+(X)<\yu(X)$  for $|X|\in\left(\frac{\pi}{2},\pi\right]$. Recalling \eqref{Y1+}, and taking the signs of $\gamma_1$ and $\cdef$ into account, we observe that $Y_1^+$ takes its maximum value  in the interval $(-\frac{\pi}{2},\frac{\pi}{2})$ at $X_0=0.$  Thus, for $|X|<\frac{\pi}{2}$, we know that $\yl {(X)}<Y_1^+(0)$. Combining the last inequality with \eqref{Y1+} and \eqref{A>1-in}, we have
\begin{align*}
\yl(X)<Y_1^+(0)&=\lA+\arsinh z_2(0)\\&<\coth \arsinh z_2(0)-\frac{k\cdef}{\gamma_1}
<-\frac{k\cdef}{\gamma_1},\quad\text{for}\quad|X|<\frac{\pi}{2}.
\end{align*} Hence $\yl(X)$ remains below $Y_1=-\frac{k\cdef}{\gamma_1}$. We set $\ym(\frac{\pi}{2})=-\frac{k\cdef}{\gamma_1}$, ensuring that $\ym$ is well-defined for all $X\in[-\pi,\pi]$.
 From \eqref{Y1-Impl}, we have 
 \begin{equation}\label{A>1-Impl}
M_1 \sin X f( \ym (X) )=\frac{\partial \ym }{\partial X} \left(  M_1 \cos Xg\!\left(\ym(X)\right)-\gamma_1\right)\!,
 \end{equation}
  for $X\in [-\pi, \pi]$. It follows that $\ym(X)$ is increasing for $X\in(0,\pi)$, and $\ym(X)$ is decreasing for $X\in(-\pi,0)$: hence  $\ym(X)$ attains its minimum at $X=0$, and its maximum at $X=\pm \pi$.

 Bearing in mind the nature of the $0-$isocline for \eqref{Upper-DS}, there are singular  points at $Q_{0}^{s}=(0,{\yl (0)})$, $Q_{0}^{c}=(0,\ym(0))$, $Q_{\pm \pi}^{s,l}=(\pm \pi,\ym(\pm \pi))$ and $Q_{\pm \pi}^{s,u}=(\pm \pi,\yu(\pm \pi))$: we claim that no part of the $\infty-$isocline intersects the line $Y_1=\lA$. 
  For $|X|<\frac {\pi} 2$, we have  $\yl (X)< \ym(X)<\lA$, and relation \eqref{A>1-in1-in} implies that
\begin{equation}\label{Prop-A_3_2-eq2}
    \ym(X)< \ym\left(\pm \frac{\pi}{2}\right)=-\frac{k\cdef}{\gamma_1}<\lA,\quad\text{for all}\quad |X|<\frac{\pi}{2}.
\end{equation}
For $|X|\in\left(\frac{\pi}{2},\pi\right]$, $\yu(X)>Y_1^+(X)>\lA$, where the last inequality follows from \eqref{Y1+}.
 It remains to show that $\ym (X)<\lA$ for $|X|\in\left(\frac {\pi} 2,\pi\right]$. Condition \eqref{A>1-in1} implies that $\ym (\pi)<\lA$, and the result follows since we have previously established that $\ym(X)$ is an increasing function on  $|X|\in\left(\frac {\pi} 2,\pi\right]$.
 
To determine the nature of the singular points, at  $Q_{\pi}^{s,u}=(\pi,\yu(\pi))$ we have 
{\begin{align*}
&D^2 H_1 (\pi,\yu(\pi))=\begin{pmatrix}
&M_1 g( \yu(\pi) ) &0\\
&0&-M_1  g( \yu (\pi) ) -\gamma_1
\end{pmatrix},
\end{align*}}
where the first term in the diagonal is negative, since $ \lA <\yu (\pi)$ and {$M_1<0$}, while the second diagonal term is positive since $\frac{\partial {F} }{\partial Y_1}(\pi,\yu(\pi))>0$. Hence Morse theory implies that $Q_{\pm\pi}^{s,u}=(\pm \pi, \yu(\pm\pi))$ are saddle points.
For $Q_{\pi}^{s,l}=(\pi,\ym(\pi))$, we get
\begin{align*}
&D^2 H_1 (\pi,\ym(\pi))=\begin{pmatrix}
&M_1 {g ( \ym(\pi))} &0\\
&0&-M_1  {g ( \ym(\pi))} -\gamma_1
\end{pmatrix},
\end{align*}
and we note that the first term on the diagonal is positive, since ${F} (X,\cdot)$ is convex, ${M_1<0}$ and $\ym(\pi)<\lA$, while the second diagonal term is negative since ${F}$ is decreasing at $(\pi,\ym(\pi))$. Thus, $Q_{\pm \pi}^{s,l}=(\pm \pi,\ym(\pm \pi))$ are saddle points as well.
For $Q_{0}^{c}=(0,\ym(0))$,
 \begin{align*}
&D^2 H_1 (0,\ym(0)) = \begin{pmatrix}
&-M_1 {g ( \ym(0))} &0\\
&0&M_1  {g ( \ym(0))} -\gamma_1
\end{pmatrix},
\end{align*}
 the first diagonal term is negative since $\ym(0)<\lA$ and ${F} (0,\cdot)$ is concave, while the second diagonal term  is also negative since $Y_1^+(0)<\ym(0)$ and ${F}$ is decreasing at $(0,\ym(0))$. Thus, Morse theory implies that $Q_{0}^{c}=(0,\ym(0))$ is a centre. 

For   $Q_{0}^{s}=(0,{\yl (0)})$ then \begin{align*}
&D^2 H_1 (0,\yl(0)) = \begin{pmatrix}
&-M_1 {g ( \yl(0))} &0\\
&0&M_1  {g ( \yl(0))} -\gamma_1
\end{pmatrix},
\end{align*}
then the first diagonal term is negative since $\yl(0)<\lA$, while $\yl(0)<Y_1^+(0)$ implies that the second diagonal term is positive, hence $Q_{0}^{s}=(0,\yl(0))$ is a saddle point.   
    \end{proof}

\begin{remark}
       The saddle point $Q_{0}^{s}=(0,\yl(0))$ which lies below the streamline of the surface wave  (in terms of $(X,Y_1)$ coordinates) is equivalent to  ${F (0,-\e_1)>0}$, due to the monotonicity of ${F} (0,\cdot)$ and, similarly, the saddle points $Q_{\pm\pi}^{s,u}=(\pm \pi,\yu(\pm \pi))$ lying above the streamline of the internal wave in $(X,Y_1)$ coordinates is equivalent to $F(\pm \pi,kh_1+\e)<0$.
    \end{remark}
    
     \begin{remark}
     In the case where the inequality \eqref{A>1-in1} does not hold then,  for some $X_0\in\left(\frac{\pi}{2},\pi\right)$, we have
 \[{F} \left(X_0,\lA\right)=0\iff M_1 \cos X_0 \sqrt{A^2-1} =k\cdef+ \gamma_1 \lA.
 \]
It follows that there are $8$ singular points for system \eqref{Upper-DS}.  These additional points are located at $(\pm X_0,\lA)$ which lie at the intersection of the branch of the $\infty-$isocline given by the graph of $\ym$, and the horizontal part of the $0-$isocline given by $Y_1=\lA.$ It can be easily verified that the Hessian matrices \eqref{Hessian_H1_fg} for $H_1$, evaluated at these points, have zero eigenvalues.  \end{remark}


\section{Discussions and conclusions}
In this paper we employed a phase plane analysis to determine qualitative features of the underlying flow for coupled periodic surface and internal waves, in a two-layer fluid model with  (constant) vorticity. The inclusion of rotational effects in the upper fluid layer represents a significant departure from the recent analyses in \cite{HV-JDE,HV-AN}, with the vorticity $\gamma_1\neq0$ complicating all aspects of the resulting wave field kinematics.  Although the waves being analysed are linear in the fluid dynamics sense --- that is, they are of (relatively) small amplitude and steepness --- the resulting dynamical systems \eqref{LowSys1} and \eqref{Upper-DS0}  which govern the motion are themselves fully nonlinear. 

We have shown that, similarly to \cite{HV-JDE,HV-AN}, the value of the non-dimensional parameter $A$ plays an overarching role in the configuration of the flow dynamics and, in particular the number, and location, of singular points. {However, unlike for the irrotational case (cf. \cite{HIV}), the inclusion of vorticity permits the possibility that $A$ is less than $1$, and indeed it may attain negative values. Disparate qualitative behaviour is observed for the three regimes: $|A|<1$, $A=1$, $A>1$, respectively.} In each of these regimes we show that three qualitatively different flow formulations may arise, depending on whether there are critical layers in the flow interior (for $\cdef=c-\gamma_1h_1<0$), or not (when $\cdef>0$), with the latter scenario having two different possible flow configurations depending on whether the vorticity $\gamma_1$ is positive, or negative.

In this paper we construct detailed qualitative depictions of the streamlines for the different possible flows in the steady moving frame. In so doing, we are undertaking an analysis of the flow dynamics from the Eulerian perspective of a fixed-observer. A natural follow-on to this will be to use these results in order to obtain a qualitative understanding of individual particle motion, as was done for the much simpler irrotational setting in \cite{HV-JDE,HV-AN}. Translating the results of this paper to such a Lagrangian framework for rotational flows is currently work in progress. Additionally, we have set the vorticity to zero in the bottom fluid layer for the purposes of expediency, and since a rotational upper layer is of greater interest both mathematically and physically. Recent work in \cite{CI-15,CI,CIM} has constructed Hamiltonian frameworks describing the evolution of internal waves in the two fluid layer model where both fluid layers possess a sheared current. An interesting extension of this paper would be to consider the underlying flow for coupled surface and internal waves whereby the lower fluid layer possesses (constant) vorticity.

\appendix

\section*{Appendix: Limiting values of the parameter $A$}\label{AppA}
In the recent paper \cite{HIV} it is proven that $A=A(h,h_1,k)\geq 1$ in the irrotational setting ($\gamma_1=0$). This result follows from determining the asymptotic values of $A$ with respect to $k$ (without loss of generality), and establishing montonicity properties for $A(k)$. 
In Remark \ref{A-pos-rem} and Lemma \ref{ALemma} we stated that richer, and more complex, behaviour is exhibited by the parameter $A=A(h,h_1,k,\gamma_1)$ in the rotational setting $(\gamma_1=0)$, as we now discuss.

For the purpose of illustration, and simplicity, we consider the behaviour of $A=A(k)$ as it varies with respect to $k$ alone. Firstly, we recall that
\begin{equation} \label{AA}
    A(k)=\frac{g-\gamma_1[ c(k)-\gamma_1 h_1 ] }{k [c(k)-\gamma_1 h_1]^2}=\frac{g-\gamma_1 \cdef (k) }{k \cdef (k) ^2 }, \quad \cdef(k):=c(k)-\gamma_1 h_1,
\end{equation}
where $c(k)$ is given by the solution of the fourth order dispersion relation \eqref{PressInt2}.
One can check that the solutions $c(k)$ are even functions of $k$, and thus $A(k)$ is an odd function of $k$: hence we consider only positive values of $k.$

To investigate asymptotic behaviour in the short-wave limit, $\lim_{k \to \infty} A(k)$, we note  that $\tanh(kh)\to 1$ exponentially fast, while terms of the form $1/(kh_1)$ have a polynomial decay of the order $-1$. Replacing $\tanh(kh) $ and $\tanh(kh_1) $ by $1$ in  \eqref{PressInt2}, this relation can then be explicitly solved to get asymptotic short-wave expressions for $c(k)$:
\begin{align}
    c_{1,2}(k)& \simeq \gamma_1 h_1 + \frac{-\gamma_1 \pm \sqrt{4 g k + \gamma_1^2} }{2k},  
    \label{asy1} \\
    c_{3,4}(k)& \simeq \gamma_1 h_1+ \frac{-2\gamma_1 h_1 k (r+2)   + \gamma_1 \pm \sqrt{4 g k r(r+2) + \gamma_1^2}}{2k(2 + r)}. \label{asy2}
\end{align}
The solutions $c_{1,2}$ correspond to barotropic (surface) wave modes, whereas $c_{3,4}$ correspond to the baroclinic (internal) wave modes. Then, from \eqref{AA} with \eqref{asy1} and \eqref{asy2} one gets
\begin{align}
  \lim_{k \to \infty}  A_{1,2}(k) =1, \quad    \lim_{k \to \infty}  A_{3,4}(k) =0.
  \end{align} 
Hence both $A(k)>1$ and $A(k)\le 1$ are possible when $\gamma_1 \ne 0$. This contrasts with the irrotational case ($\gamma_1 = 0$)  whereby we obtain the familiar asymptotic expressions (cf. \cite{Suth}) 
\begin{align*}
    c_{1,2}(k) \simeq   \pm \sqrt{ \frac{g}{k}}, \quad
    c_{3,4}(k) \simeq  \pm \sqrt{ \frac{gr} {k(2 + r)}},
\end{align*}
and
$\lim_{k \to \infty}  A_{1,2}(k) =1$, and $\lim_{k \to \infty}  A_{3,4}(k)=\frac{g }{k c_{3,4}^2 } =  1+\frac{2}{r} >1$. 
 Since one can prove in the irrotational case that the solutions $A(k)$  decrease monotonically from $\infty$ to the above limits \cite{HIV}, it follows that $A(k)\ge 1$ always. 
 
In the long-wave limit $k \to 0^+$, the wave speeds are the roots of the equation
\begin{align}\label{long_lim}
c^4 - \gamma_1 h_1 c^3 - g(h + h_1) c^2 + g \gamma_1  h h_1 c + \frac{g^2 r h h_1}{1+r}=0.
\end{align}
It follows from \eqref{AA} (see Lemma \ref{ALemma}) that $\cdef$ (and hence $c$) is real-valued  when $A\geq -\frac{\gamma_1^2}{4gk}\stackrel{k\to0}{\longrightarrow}-\infty$. Since  \eqref{long_lim} has constant real coefficients,  its roots will have finite constant values, say $c(0)$, and so  $\lim_{k \to 0^+}A(k)=  \pm \infty$  for any solution of \eqref{long_lim}, where the  sign matches that of $\text{sgn}[g+\gamma_1 ^2h_1-\gamma_1 c(0)]$. This  can be either positive or negative: for large values of $h$, the surface modes $c\approx \pm \sqrt{g(h+h_1)}$  take values larger than $\frac{g}{\gamma_1}+\gamma_1 h_1$ for
$$h \gtrsim h_0 := \frac{(g+\gamma_1^2 h_1)^2}{g \gamma_1^2}-h_1=h_1+\frac{g^2+\gamma_1^4 h_1^2}{g \gamma_1^2} \geq h_1 + 2h_1= 3h_1. $$
A degenerate case exists for $c(0)=\gamma_1 h_1$, which is possible if and only if $$\gamma_1 = \pm \frac{\sqrt{(1 + r) g r h}}{(1 + r) h_1},$$ in which case the limit is clearly
$\lim_{k \to 0^+}A(k)=\infty$.

\subsection*{Acknowledgements} 
 This publication has resulted from research conducted with financial support for DH, RI and ZS from {\em Research Ireland} under grant number 21/FFP-A/9150.

\bibliography{DH-RI-ZNS_}
\bibliographystyle{plain}

\label{lastpage}
\end{document}